\documentclass{article}

\usepackage{arxiv}

\usepackage[utf8]{inputenc} 
\usepackage[T1]{fontenc}    
\RequirePackage{amsthm,amsmath,amsfonts,amssymb}
\RequirePackage[numbers]{natbib}

\RequirePackage[colorlinks,citecolor=blue,urlcolor=blue]{hyperref}
\RequirePackage{graphicx}
\RequirePackage{tikz}
\usepackage[numbers]{natbib}

\RequirePackage{csquotes}
\usepackage{bbm}
\usepackage{mathtools}
\usepackage[normalem]{ulem}
\usepackage{xcolor}
\usepackage{multirow}
\usepackage{cancel}
\graphicspath{ {./images/} }

\DeclareMathOperator*{\argmin}{arg\,min}

\theoremstyle{plain}

\newtheorem{theorem}{Theorem}[section]
\newtheorem{corol}[theorem]{Corollary}
\newtheorem{prop}[theorem]{Proposition}
\newtheorem{lemma}[theorem]{Lemma}
\theoremstyle{definition}
\newtheorem{definition}[theorem]{Definition}

\newtheorem*{remark}{Remark}

\title{The Online Closure Principle}

\author{
 Lasse Fischer \\
  Competence Center for Clinical Trials Bremen \\ University of Bremen \\
  \texttt{fischer1@uni-bremen.de} \\
   \And
 Marta Bofill Roig \\
  Center for Medical Statistics, Informatics and Intelligent Systems \\ Medical University of Vienna \\
  \texttt{marta.bofillroig@meduniwien.ac.at} \\
  \And
 Werner Brannath \\
  Competence Center for Clinical Trials Bremen \\ University of Bremen \\
  \texttt{brannath@uni-bremen.de} \\
}

\begin{document}
\maketitle
\begin{abstract}
The closure principle is fundamental in multiple testing and has been used to derive many efficient procedures with familywise error rate control. However, it is often unsuitable for modern research, which involves flexible multiple testing settings where not all hypotheses are known at the beginning of the evaluation. In this paper, we focus on online multiple testing where a possibly infinite sequence of hypotheses is tested over time. At each step, it must be decided on the current hypothesis without having any information about the hypotheses that have not been tested yet. Our main contribution is a general and stringent mathematical definition of online multiple testing and a new online closure principle which ensures that the resulting closed procedure can be applied in the online setting. We prove that any familywise error rate controlling online procedure can be derived by this online closure principle and provide admissibility results. In addition, we demonstrate how short-cuts of these online closed procedures can be obtained under a suitable consonance property. 
\end{abstract}

\keywords{online multiple testing \and closure principle \and familywise error rate}

\section{Introduction}
 The closure principle by Marcus, Peritz and Gabriel \cite{MPG}  is one of the most fundamental principles in multiple testing, especially when considering familywise error rate (FWER) control. It has been used to derive many popular and efficient multiple testing procedures commonly applied in current practice. For example, gatekeeping procedures \cite{DOW} and graphical approaches \cite{BWBP}. Indeed, it can be shown that every FWER controlling procedure is also a closed procedure \cite{SF}. Furthermore, the closure principle can often be used to improve existing procedures \cite{GH}. In order to apply the closure principle, one needs to define the set of all non-empty intersection hypotheses, also called closure set. The idea is then to enforce coherence \cite{G} by rejecting an individual hypothesis, if all intersection hypotheses containing this individual hypothesis are rejected, each at level $\alpha$. However, 
  many modern applications do not require that all individual hypotheses are known at the beginning of the evaluation. In such cases, we must decide on individual hypotheses without knowing the intersection hypotheses formed with hypotheses added later. This makes the use of the closure principle in the classical sense impossible. One can also think the other way around. Suppose there exists an intersection test for each intersection hypothesis in the full closure set. Which properties must these intersection tests have such that we can decide on a hypothesis $H$ at a given time with the information that is available then? This is the main question we seek to answer in this paper.
 
 In this paper, we focus on the online multiple testing setting, introduced by Foster and Stine \cite{FS}. In this setting, the hypotheses arrive sequentially over time and, at each step, it must be decided whether to reject the current hypothesis without having any information about the future ones. Large internet companies, for instance, face this problem when they perform A/B tests during their marketing research \cite{Ketal}. But also in genetics, thousands of tests are carried out in a sequential manner \cite{mouse}. Javanmard and Montanari \cite{JM} even interpret scientific research itself as an online multiple testing problem, since a stream of hypotheses is continuously tested \cite{I}.  
 
Since online multiple testing was introduced in \cite{FS}, a diverse range of online procedures has been proposed \cite{JM, RYWJ, RZWJ, TR2, TR, zehetmayer2021online}. Most of these procedures provide false discovery rate (FDR) control \cite{JM, RYWJ, RZWJ, TR2, zehetmayer2021online}, where FDR is the expected proportion of true null hypotheses among the rejected hypotheses. Since FDR is less conservative than FWER \cite{BH}, it is especially useful when testing a large number of hypotheses. This is also the reason why the literature on online multiple testing has been initially focused on $\text{FDR}$ control. However, in classical applications, $\text{FWER}$ is a very common error rate and there are also online problems where it is necessary to ensure that the probability of committing any type I error is below a  certain level. For example, this may be the case in platform trials \cite{Retal} and the sequential modification of machine learning algorithms \cite{Fetal}. The paper by Tian and Ramdas \cite{TR} is the only one fully focused on online control of the FWER so far. They have introduced online versions of popular multiple testing procedures such as the graphical approach by Bretz et al. \cite{BWBP}. However, their most promising method is the ADDIS-Spending which stands for Adaptive Discarding and combines two powerful concepts used in multiple testing. First, one adapts to the number of false hypotheses, as false hypotheses cannot lead to a type I error \citep{schweder1982plots}. Second, one ignores (\enquote{discards}) hypotheses with large $p$-values such that the remaining ones can be rejected with higher probability \citep{ZSS}. In addition, online FWER control was considered in \cite{DMR} and \cite{Retal}. Döhler et al. (2021) \cite{DMR} introduced super-uniformity reward (SURE) as an alternative to discarding which is based on a priori information about the marginal CDF of null $p$-values, whereas Robertson et al. (2022) \cite{Retal} describe how to apply online error control in the context of platform trials. 
 
 Our main contribution is a novel online closure principle which ensures that the resulting closed procedure can be applied in the online setting. Note that Tian and Ramdas \cite{TR} have already provided an initial attempt for an online extension of the closure principle. However, they did not prove that the resulting closed procedure is indeed an online procedure and have therefore formulated this as an open problem. In Section \ref{sec:alpha-adj}, we show that their approach is actually a special case of a more general online closure principle.
 
 The paper begins with a general and precise definition of online multiple testing, which to the best of our knowledge, has not been introduced in the literature yet (Section \ref{sec:setup}). Afterwards, we introduce the online closure principle including a so-called predictability condition under which a closed procedure is indeed an online procedure (Section \ref{sec:online closure principle}). Moreover, we show that every FWER controlling online procedure can be obtained by this online closure principle and provide admissibility results for online closed procedures (Section \ref{sec:admissibility}). After that, we derive short-cuts of online closed procedures under consonance (Section \ref{sec:short-cuts}). In Section \ref{sec:alpha-adj}, we transfer these general results to a more specific setting in which it is assumed that a $p$-value is obtained for each individual hypothesis. This is the usual setting considered in online multiple testing literature \cite{FS, JM}. We then use the online closure principle to derive new online procedures. Particularly, this gives a uniform improvement of the currently most promising online procedure with FWER control, the ADDIS-Spending under local dependence \cite{TR}. We exemplify the usage of the proposed procedure by applying them to simulated data (Section \ref{sec:sim}) and real data of a large-scale genetic study (Section \ref{sec:IMPC}) and with an ongoing platform trial (Section \ref{sec:RECOVERY}). The paper ends with a discussion in Section \ref{sec:discussion}.
  
\section{Online multiple testing\label{sec:setup}}
In the literature \textit{online multiple testing} is described as the setting, where an infinite stream of null hypotheses $\mathcal{H}=(H_i)_{i\in \mathbb{N}}$, indexed by entry order, is tested in a sequential manner. This means that at each step/time $i\in \mathbb{N}$ it needs to be decided whether $H_i$ is rejected without access to any information about the future hypotheses and data \cite{FS, JM}. In this section, we define online multiple testing in a more mathematical manner so that it becomes clearer what a multiple testing procedure must satisfy to be termed online.

Let $(\Omega,\mathcal{A})$ be a measurable space and $\mathcal{P}$ some set of probability distributions on $(\Omega, \mathcal{A})$. Note that $(\Omega, \mathcal{A}, \mathcal{P})$ is to be understood in an abstract sense and it is not supposed to be completely known in advance. We assume that the data  follows some unknown distribution $\mathbb{P}\in \mathcal{P}$.
 The hypotheses $(H_i)_{i \in \mathbb{N}}$ can be formally considered as subsets of $\mathcal{P}$ and by testing $H_i$, we want to examine whether $\mathbb{P} \in H_i$. Unless otherwise stated, equalities and inequalities involving random variables should be understood to hold almost surely for all $\mathbb{P} \in \mathcal{P}$. We further assume that a filtration $\mathbb{F}=(\mathcal{F}_i)_{i\in \mathbb{N}}$ (increasing sequence of $\sigma$-fields) is given, where $\mathcal{F}_i\subseteq \mathcal{A}$ defines the information that the test decision for $H_i$ is allowed to depend on. For example, $\mathcal{F}_i$ can be the $\sigma$-field that is generated by all observations that are available at time $i$. However, we may do not want to use all observations completely or add external randomization. With this, we can formally define an online multiple testing procedure as follows.
 
\begin{definition}[Online multiple testing procedure\label{def:online multiple testing}]
An \textit{online multiple testing procedure} (hereinafter referred to as online procedure for short) for $\mathcal{H}=(H_i)_{i\in \mathbb{N}}$ is a sequence of test decisions $\boldsymbol{d}=(d_i)_{i\in \mathbb{N}}$, where each $d_i$ is a random variable with values in $\{0,1\}$ that is measurable with respect to $\mathcal{F}_i$. If $d_i=1$, we conclude that $H_i$ is rejected and if $d_i=0$, that $H_i$ is accepted.
\end{definition}
Therefore, in contrast to classical \enquote{offline} multiple testing, we have an infinite number of hypotheses to consider. Furthermore, each test decision $d_i$ is only allowed to use some partial information $\mathcal{F}_i$ of the total information $\mathcal{A}$, whereby the partial information $\mathcal{F}_i$ is growing over time $i\in \mathbb{N}$. 
Note that this setting encompasses the classical setting as a special case, in which $\mathcal{F}_i=\mathcal{A}$ for all $i\in \mathbb{N}$ and the testing process is stopped after $m \in \mathbb{N}$ steps, e.g. by choosing $H_i=\mathcal{P}$ and $d_i=0$ for all $i>m$. In a similar way, \textit{online batch testing} \citep{zrnic2020power} can be embedded in our framework. Even though the theoretical results of this paper apply in this general setting, we focus on the strict online case of $\mathcal{F}_1 \subset \mathcal{F}_2 \subset ...$  when deriving concrete online closed procedures.

We denote by $I_0^\mathbb{P}\coloneqq \{i\in \mathbb{N}: \mathbb{P} \in H_i\}$ and $I_1^\mathbb{P}\coloneqq \mathbb{N}\setminus I_0^\mathbb{P}$ the index sets of true and false null hypotheses, if $\mathbb{P}$ was the true distribution, respectively. Furthermore, for all $i\in \mathbb{N}$, we define $v_{\mathbb{P}}(i)\coloneqq \sum_{j\leq i, j\in I_0^\mathbb{P}} d_j$ as the number of falsely rejected hypotheses up to step $i\in \mathbb{N}$ and set $v_{\mathbb{P}}\coloneqq \lim\limits_{i\rightarrow \infty}v_{\mathbb{P}}(i)$. With this, we define the \textit{familywise error rate} (FWER) at time $i\in \mathbb{N}$ and over all hypotheses as
\begin{align}
    \text{FWER}_{\mathbb{P}}(i)\coloneqq \mathbb{P}(v_{\mathbb{P}}(i)>0) \quad \text{ and } \quad \text{FWER}_{\mathbb{P}}\coloneqq \mathbb{P}(v_{\mathbb{P}}>0) \qquad (\mathbb{P} \in \mathcal{P}).
\end{align}
We aim for strong control of the FWER at each time $i\in \mathbb{N}$, which means that for some pre-specified $\alpha\in (0,1)$, we have $\text{FWER}_{\mathbb{P}}(i)\leq \alpha$ for all $i\in \mathbb{N}$ and $\mathbb{P}\in \mathcal{P}$. 
Note that this is equivalent to requiring $\text{FWER}_{\mathbb{P}}\leq \alpha$ for all $\mathbb{P}\in \mathcal{P}$, since $E_i^{\mathbb{P}}\coloneqq \{v_{\mathbb{P}}(i)>0\}\in \mathcal{F}_i$ is an increasing sequence ($E_1^{\mathbb{P}} \subseteq E_2^{\mathbb{P}} \subseteq  \ldots$) with $E_i^{\mathbb{P}}\subseteq \{v_{\mathbb{P}}>0\}$ for all $i\in \mathbb{N}$. Therefore, we drop the index $i$ in the following. In contrast to strong control, weak FWER control only requires that $\text{FWER}_{\mathbb{P}}\leq \alpha$ for all distributions contained in the global null hypothesis $\mathbb{P} \in \bigcap_{i\in \mathbb{N}} H_i$. This is only of limited use in practice. Hence, when we write control in the remainder of this paper, we always mean strong control.

\section{Online closure principle\label{sec:online closure principle}}

For a potentially infinite index set $I\subseteq \mathbb{N}$, we denote  the corresponding intersection hypothesis and intersection test by $H_I=\bigcap_{i\in I}H_i$ and $\phi_I$, respectively. Each $\phi_I$ is a random variable with values in $\{0,1\}$ such that $H_I$ is rejected by $\phi_I$, if $\phi_I=1$, and accepted, if $\phi_I=0$. We say that $\phi_I$, $I\subseteq \mathbb{N}$, is an online intersection test, if $\phi_I$ is measurable with respect to $\mathcal{F}_{\sup(I)}$, where $\mathcal{F}_{\infty}=\mathcal{A}$.
 Furthermore, $\phi_I$ is an $\alpha$-level intersection test, if $\mathbb{P}(\phi_I=1)\leq \alpha$ for all $\mathbb{P}\in H_I$. For the online closure principle, we need an online $\alpha$-level intersection test $\phi_I$ for each $I\subseteq \mathbb{N}$, where we always set $\phi_{\emptyset}=0$. We will see that if the family $\boldsymbol{\phi}=(\phi_I)_{I\subseteq \mathbb{N}}$ fulfils the following condition, the resulting closed testing procedure is indeed an online procedure.


\begin{definition}[Predictable family of online intersection tests\label{def: predictable family}]
A family of online intersection tests $\boldsymbol{\phi}=(\phi_I)_{I\subseteq \mathbb{N}}$ is called \textit{predictable}, if for all $i\in \mathbb{N}$ and $I\subseteq \{1,\ldots,i\}$ holds that: 
 $$\phi_I=1 \text{ implies } \phi_K=1 \text{ for all } K= I \cup J \text{ with } J\subseteq \{j\in \mathbb{N}: j>i\}.$$
\end{definition}
This predictability condition ensures that if a finite intersection hypothesis $H_I$, $I\subseteq \{1,\ldots,i\}$, is rejected, it remains rejected when future hypotheses $H_j$, $j>i$, are added. For example, suppose $H_1\cap H_3$ is rejected, then $H_1\cap H_3 \cap H_4$ needs to be automatically rejected as well. However, $H_1\cap H_2 \cap H_3$ does not need to be rejected, as $H_2$ is not a future hypothesis in that case. Now, we can formulate a closure principle for online multiple testing.

\begin{theorem}[Online closure principle\label{theo:online closure principle}]
Let $\boldsymbol{\phi}=(\phi_I)_{I \subseteq \mathbb{N}}$ be an arbitrary family of $\alpha$-level intersection tests. Then, the closed procedure $\boldsymbol{d}^{\boldsymbol{\phi}}=(d_i^{\boldsymbol{\phi}})_{i\in \mathbb{N}}$ based on $\boldsymbol{\phi}$ defined by
\begin{eqnarray*}
    d_i^{\boldsymbol{\phi}} &=& \min\{\phi_I : I \subseteq \mathbb{N} \text{ with } i \in I\}
\end{eqnarray*}
controls the FWER at level $\alpha$ in the strong sense.
In addition, if each $\phi_I$ is an online intersection test and the family of online intersection tests $\boldsymbol{\phi}$ is predictable, then $\boldsymbol{d}^{\boldsymbol{\phi}}$ is an online procedure. We refer to such procedures as \textit{online closed procedures}.
\end{theorem}

To prove Theorem \ref{theo:online closure principle}, we first show the following lemma, which states that if the predictability condition is satisfied, only the current and previous hypotheses need to be considered at each step.
\begin{lemma}\label{lemma:closure principle}
 If $\boldsymbol{\phi}=(\phi_I)_{I \subseteq \mathbb{N}}$ is predictable, then $d_i^{\boldsymbol{\phi}}$ defined in Theorem 3.2 satisfies $d_i^{\boldsymbol{\phi}}= \min\{\phi_I: I \subseteq \{1,\ldots,i\} \text{ with } i \in I\}$ for all $i\in \mathbb{N}$. 
\end{lemma}
\begin{proof}
Let $i\in \mathbb{N}$ and $K\subseteq \mathbb{N}$ with $i\in K$ be arbitrary. Note that $K$ can be written as $K=I \cup J$, where $I=\{k\in K: k\leq i\}$ and $J=\{k\in K: k>i\}$. The predictability of $(\phi_I)_{I\subseteq \mathbb{N}}$ ensures that $H_K$ is rejected by $\phi_K$ if $H_I$ is rejected by $\phi_I$. 
\end{proof}

\begin{proof}[Proof of Theorem \ref{theo:online closure principle}]
We first show FWER control. Let $\mathbb{P}\in \mathcal{P}$ be arbitrary. In order to reject any true null hypothesis, it needs to hold for the subset containing the indices of all true hypotheses $I_0^{\mathbb{P}}$ that $\phi_{I_{0}^{\mathbb{P}}}=1$. Since $\phi_{I_{0}^{\mathbb{P}}}$ is an  $\alpha$-level intersection test, we have $$\mathbb{P}(v_{\mathbb{P}}> 0) \leq \mathbb{P}(\phi_{I_{0}^{\mathbb{P}}}=1) \leq \alpha.$$
To show the second assertion, we assume that each $\phi_I$ is an online intersection test and
$\boldsymbol{\phi}=(\phi_I)_{I\subseteq \mathbb{N}}$ is predictable. Lemma \ref{lemma:closure principle} implies that $d_i^{\boldsymbol{\phi}}= \min\{\phi_I : I \subseteq \{1,\ldots,i\} \text{ with } i \in I\}$, $i\in \mathbb{N}$. Since each $\phi_I$ with $I\subseteq \{1,\ldots,i\}$ is measurable with respect to $\mathcal{F}_{\sup(I)}\subseteq \mathcal{F}_{i}$, $d_i^{\boldsymbol{\phi}}$ is measurable with respect to $\mathcal{F}_i$. 
\end{proof}


Note that Lemma \ref{lemma:closure principle} does not hold in general when the predictability of $(\phi_I)_{I\subseteq \mathbb{N}}$ is violated. For example, let $p_1$ and $p_2$ be the $p$-values for $H_1$ and $H_2$ that are measurable with respect to $\mathcal{F}_1$ and $\mathcal{F}_2\supset \mathcal{F}_1$, respectively. Suppose $(\phi_I)_{I\subseteq \mathbb{N}}$ is a family of online intersection tests such that $\phi_{\{1\}}=1$ if $p_1 \leq \alpha$ and $\phi_{\{1,2\}}=1$ if $p_1 \leq \frac{\alpha}{2}$ or $p_2 \leq \frac{\alpha}{2}$. Now assume that $\frac{\alpha}{2}<p_1\leq \alpha$ and $p_2>\frac{\alpha}{2}$. Thus, $\phi_{\{1\}}=1$ but $\phi_{\{1,2\}}=0$ and rejecting $H_{\{1\}}$ by $\phi_{\{1\}}$ would not be sufficient to reject $H_1$ by the closure principle. This implies that this closed procedure cannot be an online procedure and that the assumption of $\phi_I$, $I\subseteq \mathbb{N}$, being an online intersection test is insufficient to obtain an online closed procedure. In the following section, we show that predictability is necessary in general.

\subsection{Admissibility of online closed procedures\label{sec:admissibility}}
There is a large body of literature discussing the admissibility of classical closed procedures \citep{GHS, romano2011consonance, lehmann1986testing, SF}. Sonnemann and Finner (1988) \citep{SF} showed that every admissible procedure with FWER control can be derived as a closed procedure and Romano et al. (2011) \citep{romano2011consonance} proved that one can further restrict to consonant intersection tests. In this section, we derive similar admissibility results for the online closure principle (Theorem \ref{theo:online closure principle}). We follow with our definition of admissibility the one in \citep{GHS}. 

 \begin{definition}[Admissibility of online procedures\label{def:admiss}]
     A strong FWER controlling online procedure is called admissible when it cannot be uniformly improved by another online procedure with strong FWER control, where $\boldsymbol{d}=(d_i)_{i\in \mathbb{N}}$  is uniformly improved by $\boldsymbol{\tilde{d}}=(\tilde{d}_i)_{i\in \mathbb{N}}$, if $\tilde{d}_i\geq d_i$ for all $i\in \mathbb{N}$ and $\mathbb{P} (\tilde{d}_i>d_i)>0$ for some $i\in \mathbb{N}$ and $\mathbb{P}\in \mathcal{P}$.
 \end{definition} 
 
In the following theorem, we prove that any online procedure with strong FWER control can be obtained by the online closure principle (Theorem \ref{theo:online closure principle}). This shows that the fundamentality of the classical closure principle can be transferred to the online setting. Furthermore, it implies that the predictability condition (Definition \ref{def: predictable family}) is not too strict.

\begin{theorem}\label{theo:closed_general}
 Let $\boldsymbol{d}=(d_i)_{i\in \mathbb{N}}$ be an online procedure with strong FWER control. Then $\boldsymbol{\phi}=(\phi_I)_{I\subseteq  \mathbb{N}}$, where $\phi_I=\max\{d_i: i\in I\}$, is a predictable family of online $\alpha$-level intersection tests and $\boldsymbol{d}^{\boldsymbol{\phi}}=\boldsymbol{d}$. Thus, for any online procedure $\boldsymbol{d}$ with FWER control there exists an online closed procedure $\boldsymbol{d}^{\boldsymbol{\phi}}$ that leads to the same decisions.
\end{theorem}
\begin{proof}
     Since $\boldsymbol{d}=(d_i)_{i\in \mathbb{N}}$ is an online procedure, $\phi_I=\max\{d_i: i\in I\}$ is measurable with respect to $\mathcal{F}_{\sup(I)}$ and thus defines an online intersection test for all $I\subseteq \mathbb{N}$. Given the strong FWER control of $\boldsymbol{d}$, it follows that $\phi_I$ is an $\alpha$-level intersection test. To see this, suppose $I\subseteq I_0^{\mathbb{P}}$ for some $\mathbb{P}\in \mathcal{P}$. Hence, $\mathbb{P}(\phi_I=1)=\mathbb{P}\left(\max\{d_i:i\in I\}= 1\right) \leq \mathbb{P}\left(\max\{d_i:i\in I_0^{\mathbb{P}}\}= 1\right)\leq \alpha$. Furthermore, $\phi_I=1$ implies $\phi_K=1$ for all $I\subseteq K$, which ensures the predictability of $\boldsymbol{\phi}=(\phi_I)_{I\subseteq \mathbb{N}}$. It remains to show that $\boldsymbol{d}^{\boldsymbol{\phi}}=\boldsymbol{d}$. First, note that $d_i=0$ implies $\phi_{\{i\}}=0$ and thus $d^{\boldsymbol{\phi}}_i=0$ for all $i\in \mathbb{N}$. Second, $d_i=1$ implies $\phi_{I}=1$ for all $I \subseteq \mathbb{N}$ with $i\in I$ and hence $d^{\boldsymbol{\phi}}_i=1$.
\end{proof}


  A family of intersection tests  $\boldsymbol{\phi}=(\phi_I)_{I\subseteq \mathbb{N}}$ is \textit{consonant} \citep{G}, if for all $I\subseteq\mathbb{N}$: \begin{align} \phi_I=1 \text{ implies } \exists i\in I: \phi_J=1 \text{ } \forall J \subseteq I \text{ with } i\in J. \label{eq:consonance}\end{align}
 If a family of intersection tests is not consonant, it is called \textit{dissonant}. Closed procedures based on consonant intersection tests have the desirable property that the rejection of an intersection hypothesis $H_I$ implies that at least one individual hypothesis $H_i$ with $i\in I$ is rejected. For example, suppose we are testing several treatment arms $T_1,T_2,\ldots$ against a common control (e.g. in a platform trial). Then the rejection of $H_1\cap H_2$ would imply that at least $T_1$ or $T_2$ is efficient. However, if the procedure is dissonant, we might not be able to conclude which of the two treatments is efficient. Romano et al. (2011) \citep{romano2011consonance} showed that every strong FWER controlling online procedure can be written as a closed procedure based on consonant intersection tests. Since the $\boldsymbol{\phi}$ defined in Theorem \ref{theo:closed_general} is consonant, it immediately follows that this result also applies in the online setting. 



\begin{corol}\label{corol:consonance}
For every online procedure $\boldsymbol{d}$ that controls the FWER strongly, there exists an online closed procedure $\boldsymbol{d}^{\boldsymbol{\phi}}$ with $\boldsymbol{d}^{\boldsymbol{\phi}}=\boldsymbol{d}$ that is based on a consonant family of intersection tests $\boldsymbol{\phi}$.
\end{corol}


Corollary \ref{corol:consonance} only states that any online procedure can also be written as an online closed procedure based on consonant intersection tests. However, Romano et al. (2011) \citep{romano2011consonance} have shown that \enquote{consonantizing} dissonant intersection tests $\boldsymbol{\phi}=(\phi_I)_{I\subseteq \mathbb{N}}$ by choosing $\tilde{\phi}_I=\max\{d_i^{\boldsymbol{\phi}}:i\in I\}$, $I\subseteq \mathbb{N}$, often reveals weaknesses of $\boldsymbol{d}^{\boldsymbol{\phi}}$, which can be used to improve $\boldsymbol{d}^{\boldsymbol{\phi}}$ by improving $\boldsymbol{\tilde{\phi}}$. We also illustrate this in Section \ref{sec:closed alpha-spending} by an example. Note that in the online case one needs to be careful with constructing improvements of intersection tests, as the predictability might get lost and the closed procedure is no longer an online procedure. For this reason, the following result may be helpful.

 \begin{prop}\label{prop:infinite_admissible}
 Let $\boldsymbol{\phi}=(\phi_I)_{I\subseteq \mathbb{N}}$ be a predictable family of online intersection tests. Furthermore, let $\psi_I=\phi_I$ for all finite index sets $I\subseteq \mathbb{N}$ and $\psi_I \geq \phi_I$   for all infinite $I\subseteq \mathbb{N}$. Then $\boldsymbol{\psi}=(\psi_I)_{I\subseteq \mathbb{N}}$ is a predictable family of online intersection tests as well.
 \end{prop}
 \begin{proof}
Let $I\subseteq \{1,\ldots,i\}$ for some $i\in \mathbb{N}$ and $K=I\cup J$ with $J\subseteq \{j\in \mathbb{N}: j>i\}$. By the predictability of $\boldsymbol{\phi}$, we have $\psi_I=\phi_I\leq \phi_K \leq \psi_K$, which shows the predictability of $\boldsymbol{\psi}$. Furthermore, $\psi_I$ is an online intersection test for each $I\subseteq \mathbb{N}$ by definition. 
 \end{proof}

The proposition shows that we cannot violate the predictability condition by improving intersection tests $\phi_I$ for infinite $I\subseteq \mathbb{N}$. Therefore, one approach to improve an existing online procedure $\boldsymbol{d}$ using the online closure principle would be to define $\phi_I=\max\{d_i:i\in I\}$ as in Theorem \ref{theo:closed_general}. Then, if possible, uniformly improve $\phi_I$ by $\psi_I$ for infinite $I\subseteq \mathbb{N}$. After that, one might be able to also uniformly improve $\phi_I$ by $\psi_I$ for finite $I\subseteq \mathbb{N}$ while retaining predictability of $(\psi_I)_{I\subseteq \mathbb{N}}$. Note that an improvement of some or all $\phi_I$ for infinite $I$ lead to a relaxation of the predictability condition and thereby could create possibilities to improve $\phi_I$ for finite $I$. 

For example, for all $i\in \mathbb{N}$ let $p_i$ be a $p$-value for  $H_i$ that is measurable with respect to $\mathcal{F}_i$. Then $\boldsymbol{d}=(d_i)_{i\in \mathbb{N}}$ with $d_i=\mathbbm{1}\{p_i\leq \alpha_i\}$, where $\alpha_i>0$ and $\sum_{i\in \mathbb{N}} \alpha_i=\alpha$, defines an online procedure with FWER control due to Bonferroni's inequality. Since $\sum_{i\in I} \alpha_i<\alpha$ for every $I\subset \mathbb{N}$, the intersection tests $\phi_I=\max\{d_i:i\in I\}$ can be improved. For arbitrary $p$-values and a finite $I\subseteq \mathbb{N}$ an improvement of $\phi_I$ leads to a violation of the predictability condition, however, due to Proposition \ref{prop:infinite_admissible} we can safely improve $\phi_I$ for infinite $I\subseteq \mathbb{N}$. For instance, define $\psi_I=\max\{\mathbbm{1}\{p_i\leq \alpha_i^I\}:i\in I\}$, where $\alpha_i^I\geq \alpha_i$ and $\sum_{i\in I} \alpha_i^I=\alpha$, for all infinite $I\subseteq \mathbb{N}$. Now, we can also improve $\phi_I$ for finite $I$ by $\psi_I=\max\{\mathbbm{1}\{p_i\leq \alpha_i^I\}:i\in I\}$, where $\alpha_i^I=\inf\{\alpha_i^K: \exists J\subseteq \mathbb{N}\setminus \{1,\ldots,\max(I)\}, J \text{ infinite, } K=I\cup J\}$. Then $\boldsymbol{\psi}=(\psi_I)_{I\subseteq \mathbb{N}}$ is a predictable family of online $\alpha$-level intersection tests with $\psi_I\geq \phi_I$ for all $I \subseteq \mathbb{N}$. In Section \ref{sec:closed alpha-spending} and \ref{sec:online-graph} we derive concrete improvements of this Alpha-Spending procedure \citep{FS}.

Theorem \ref{theo:closed_general} and Corollary \ref{corol:consonance} show that predictability and consonance of the family of intersection tests are necessary conditions for admissibility of an online procedure. Furthermore, Proposition \ref{prop:infinite_admissible} implies that if admissible intersection tests exist, admissible intersection tests for infinite index sets are also necessary for admissibility of an online procedure. Analogously to Definition \ref{def:admiss}, a single $\alpha$-level test $\delta$ for a hypothesis $H\subseteq \mathcal{P}$ is admissible, if there exists no other $\alpha$-level test $\tilde{\delta}$ for $H$ with $\tilde{\delta}\geq \delta$ and $\mathbb{P} (\tilde{\delta}>\delta)>0$ for some $\mathbb{P} \in \mathcal{P}$ \citep{GHS, lehmann1986testing}. But, as in classical multiple testing, it is difficult (or even impossible) to find non-trivial sufficient conditions for admissibility. For example, it is not ensured that a closed procedure $\boldsymbol{d}^{\boldsymbol{\phi}}$ is admissible, if $\phi_I$ is admissible for all $I\subseteq \mathbb{N}$ \citep{bittman2009optimal, GHS}.  As pointed out by Goeman et al. (2021) \citep{GHS} showing admissibility for multiple testing procedures can be resolved by consideration of a monotone family of procedures $(\boldsymbol{d}^I)_{I\subseteq \mathbb{N}}$ that defines a multiple testing procedure for each subset of hypotheses, which we not do in this paper. However, we can prove a condition under which the event of rejecting any hypothesis cannot be enlarged without violating FWER control. 


\begin{prop}\label{prop:optimal_rejecting_one}
    Let $\boldsymbol{\phi}=(\phi_I)_{I\subseteq \mathbb{N}}$ be a consonant family of intersection tests. If $\phi_\mathbb{N}$ is admissible, there does not exist a strong FWER controlling procedure $\boldsymbol{d}=(d_i)_{i\in \mathbb{N}}$ with $\max\{d_i: i\in \mathbb{N}\}\geq \max\{d_i^{\boldsymbol{\phi}}: i\in \mathbb{N}\}$ and $\mathbb{P}(\max\{d_i: i\in \mathbb{N}\}> \max\{d_i^{\boldsymbol{\phi}}: i\in \mathbb{N}\})>0$ for some $\mathbb{P} \in \mathcal{P}$. 
\end{prop}
\begin{proof}
    Suppose there exists a $\boldsymbol{d}$ with the property stated in the theorem and define $E\coloneqq \left\{ \max\{d_i: i\in \mathbb{N}\}> \max\{d_i^{\boldsymbol{\phi}}: i\in \mathbb{N}\} \right\}$. Further, let $\tilde{\phi}_I=\max\{d_i: i\in I\}$, $I\subseteq \mathbb{N}$, be the intersection test defined in Theorem \ref{theo:closed_general}. Since $\boldsymbol{\phi}$ is consonant, we have
    $\phi_\mathbb{N}=\max\{d_i^{\boldsymbol{\phi}}: i\in \mathbb{N}\}\leq \max\{d_i: i\in \mathbb{N}\}=\tilde{\phi}_\mathbb{N}$
    with a strict inequality if $E$ happens. This contradicts the admissibility of $\phi_\mathbb{N}$.
\end{proof}

\begin{remark}
    Proposition \ref{prop:optimal_rejecting_one} is inspired by a result shown in \cite{romano2011consonance}. They considered the case, where the global test $\phi_{\mathbb{N}}$ maximizes the minimum probability of rejecting $H_{\mathbb{N}}$ over some set of alternative distributions $H_A \subseteq \mathcal{P}\setminus H_{\mathbb{N}}$ and showed that any consonant closed procedure based on this $\phi_{\mathbb{N}}$ also maximizes the minimum probability of rejecting any hypothesis over $H_A$. We think our result fits the online setting better, since one does usually not consider a fixed set of alternatives. Furthermore, the conditions of our proposition are also easy to meet in the online case as shown by a simple example in the following. 
\end{remark}

Proposition \ref{prop:optimal_rejecting_one} is not restricted to online procedures and thus it might seem that the sufficient condition is difficult to achieve in the online case. However, we already showed that making  a predictable family of online intersection tests consonant (Corollary \ref{corol:consonance}) or uniformly improving its infinite intersection tests (Proposition \ref{prop:infinite_admissible}), will always lead to a predictable family of online intersection tests again. Therefore, it should not be too hard to meet these conditions. For example, suppose we have independent $p$-values $(p_i)_{i\in \mathbb{N}}$ that are uniformly distributed under the null hypothesis. It can then be shown that using the Online-\u{S}id\'{a}k procedure \cite{TR}, which is defined by $d_i=\mathbbm{1}\{p_i\leq \alpha_i\}$ with $\alpha_i=1-(1-\alpha)^{\gamma_i}$ and $\sum_{i\in \mathbb{N}} \gamma_i=1$, the probability of rejecting any hypothesis is exactly $\alpha$ under the global null hypothesis. Thus, if we define $\boldsymbol{\phi}$ as in Theorem \ref{theo:closed_general} for the Online-\u{S}id\'{a}k, we obtain a consonant and predictable family of online intersection tests, where the test $\phi_{\mathbb{N}}$ has exact size $\alpha$. Under mild assumptions, e.g. that the collection of null sets is the same for all distributions $\mathbb{P} \in \mathcal{P}$ \cite{GHS}, this implies that $\phi_{\mathbb{N}}$ is admissible. Thus, in this setting the event of rejecting any hypothesis by Online-\u{S}id\'{a}k cannot be enlarged without violating FWER control.

\subsection{Short-cuts under consonance\label{sec:short-cuts}}

By Theorem \ref{theo:closed_general}, we can focus completely on online closed procedures when constructing new online procedures with FWER control and, by Corollary \ref{corol:consonance}, we may even restrict to consonant intersection tests. Usually, at each step $i \in \mathbb{N}$ we need to consider up to $2^{i-1}$ intersection hypotheses $H_I$, $I\subseteq \mathbb{N}$, with $i\in I$. Since $i$ tends to infinity, it is unrealisable to test all of these intersection hypotheses in practice. Even if the testing process stops at some point, it is computationally intensive and difficult to communicate. In the offline setting, the same problems occur when a large number of hypotheses is tested, which led to the establishment of short-cut procedures \cite{GH}. The objective of short-cut procedures is to find decisions for the individual hypotheses without testing every intersection hypothesis. In this way, the number of operations should be reduced to the number of individual hypotheses while the decisions coincide with those of a closed procedure. We now want to apply this approach to the online case. To formulate a short-cut, additional assumptions towards the family of intersection tests $\phi$ are required. In this paper, we focus on short-cuts based on consonance \eqref{eq:consonance}.

When constructing consonance-based short-cuts in offline multiple testing, one would usually start with the global hypothesis \cite{HBW}, which is the intersection of all individual hypotheses. If the global hypothesis is rejected, there exists an index $i$ satisfying the consonance property \eqref{eq:consonance}, which implies that the individual hypothesis $H_i$ can be rejected by the closure principle. In the next step, the intersection of all hypotheses except $H_i$ is considered and the testing step is repeated. This can be continued until the intersection of the remaining hypotheses cannot be rejected. In online multiple testing, this proceeding is  not possible as the global hypothesis is not known at the beginning of the evaluation. However, the predictability condition makes it possible to formulate a short-cut anyway.

Assume $(\phi_I)_{I \subseteq \mathbb{N}}$ is a predictable family of online intersection tests with the consonance property and consider the intersection hypothesis $H_{\{1\}}$. When $H_{\{1\}}$ is rejected by its online intersection test $\phi_{\{1\}}$, the predictability of $(\phi_I)_{I \subseteq \mathbb{N}}$ ensures that $H_1$ is rejected by the online closure principle (Lemma \ref{lemma:closure principle}). Now, set $I_2=\{2\}$ if $\phi_{\{1\}}=1$, and $I_2=\{1,2\}$ otherwise. Suppose $\phi_{I_2}=1$. In case of $I_2=\{2\}$, it holds $\phi_{\{1\}}=1$ and due to the predictability of $(\phi_I)_{I \subseteq \mathbb{N}}$, it also holds $\phi_{\{1,2\}}=1$. Hence, $\phi_{\{1,2\}}=1$ and $\phi_{\{2\}}=1$ implying that $H_2$ is rejected by the closure principle (Lemma \ref{lemma:closure principle}). If $I_2=\{1,2\}$, the consonance property implies that $\phi_{\{2\}}=1$ as well and again $H_2$ is rejected by the closure principle. This can be continued and a short-cut of the closed procedure is obtained, meaning that only one intersection hypothesis is tested for each individual hypothesis $H_i$, $i\in \mathbb{N}$. An illustration of the short-cut can be found in Figure \ref{fig:short-cut}. A formal description is given in the next theorem, whose proof can be found in the Appendix.

\begin{figure}
\centering
\includegraphics[width=13cm]{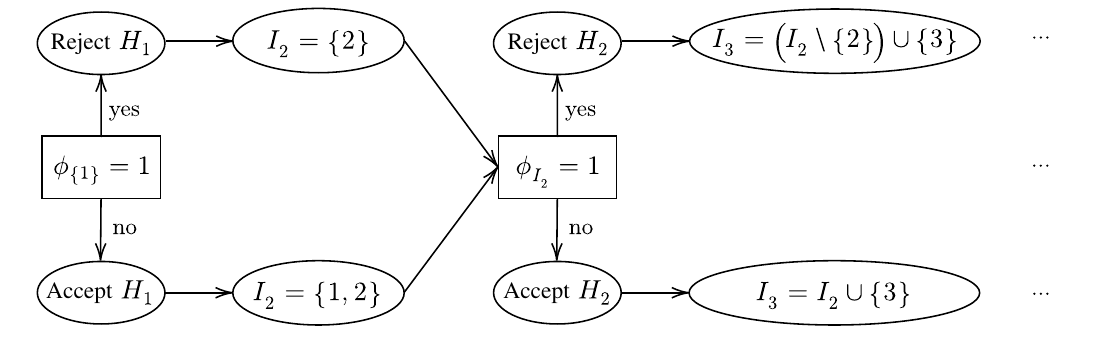}%
\caption{Short-cut of an online closed procedure based on consonance.\label{fig:short-cut}}\end{figure}

\begin{theorem}\label{theo:consonance procedure}
Assume $\boldsymbol{\phi}=(\phi_I)_{I \subseteq \mathbb{N}}$ is a predictable family of online $\alpha$-level intersection tests with the consonance property. Let us recursively define $I_1=\{1\}$ and $I_i=\{j\in \mathbb{N}: j<i, \phi_{I_j}=0\} \cup \{i\}$ for all $i\geq 2$. Then the following procedures lead to the same decisions:
\begin{enumerate}
\item The online closed procedure $\boldsymbol{d}^{\boldsymbol{\phi}}$.
\item The short-cut $\boldsymbol{d}^{\boldsymbol{\phi},s}=(d^{\boldsymbol{\phi},s}_i)_{i\in \mathbb{N}}$, where $d^{\boldsymbol{\phi},s}_i=\phi_{I_i}$ for all $i\in \mathbb{N}$.
\end{enumerate}
\end{theorem}

Note that if an intersection hypothesis $H_{I_i}$, $i\in \mathbb{N}$, is rejected, it is also uniquely determined which individual hypothesis, namely $H_i$, is to be rejected, whereas in the classical case, the index satisfying the consonance property must first be determined. In the next  section, we will use this fact to calculate individual significance levels for the short-cut. By means of such a result, we will obtain new and more  powerful online procedures. For this purpose, we consider in the next section a more specific online multiple testing setting that is based on $p$-values.

\section{Online closed testing with $\alpha$-adjustments\label{sec:alpha-adj}}

 Most existing online multiple testing procedures are defined based on $p$-values $(p_i)_{i\in \mathbb{N}}$ for the individual hypotheses $(H_i)_{i\in \mathbb{N}}$ \cite{FS, JM, TR}. Each $p$-value $p_i$ can be considered as a random variable with values in $[0,1]$ that is measurable with respect to $\mathcal{F}_i$.  It is assumed that all $p$-values are valid, which means $\mathbb{P}(p_i \leq x)\leq x$ for all $\mathbb{P}\in H_i$ and $x \in [0,1]$. Using $p$-values, a null hypothesis $H_i$ is rejected if $p_i\leq \alpha_i$, where $\alpha_i\in [0,1)$ is the individual significance level for $H_i$. We call a sequence $(\alpha_i)_{i\in \mathbb{N}}$ \textit{$\alpha$-adjustment} and the multiple testing procedure $\boldsymbol{d}=(d_i)_{i\in \mathbb{N}}$ with $d_i=\mathbbm{1}\{p_i\leq \alpha_i\}$ \textit{$\alpha$-adjustment procedure}. An $\alpha$-adjustment defines an online procedure, if $\alpha_i$ is measurable with respect to $\mathcal{F}_i$. In this case, we refer to it as \textit{online $\alpha$-adjustment}. It should be noted that in order to obtain online procedures with the required error rate control, $p_i$ and $\alpha_i$ are often chosen to be measurable with respect to smaller $\sigma$-fields $\mathcal{X}_i\subseteq \mathcal{F}_i$ and $\mathcal{Y}_i\subseteq \mathcal{F}_i$, respectively, such that $\mathbb{P}(p_i\leq \alpha_i |\mathcal{Y}_i)\leq \alpha_i$ for all $\mathbb{P}\in H_i$. However, this is contained in our more general setting. For example, the literature often considers the case where solely the $p$-values $p_1, \ldots, p_i$ are available at time $i\in \mathbb{N}$. Thus, we have $\mathcal{F}_i=\sigma(p_1,\ldots,p_i)$. The individual significance levels $(\alpha_i)_{i\in \mathbb{N}}$ are then usually chosen as non-random functions of indicators (e.g. rejections) of the previous $p$-values, which ensures that $\alpha_i$ is measurable with respect to $\mathcal{F}_{i-1}\subseteq \mathcal{F}_{i}$. If the $p$-values are independent, we have $\mathbb{P}(p_i\leq \alpha_i |\mathcal{F}_{i-1})\leq \alpha_i$ for all $\mathbb{P}\in H_i$. 

One can also use $\alpha$-adjustments to test intersection hypotheses $H_I$, $I\subseteq \mathbb{N}$.
 In this case, we only require an individual significance level $\alpha_i^I$ for each $H_i$ with $i\in I$. We define $\boldsymbol{\alpha}_I=(\alpha_i^I)_{i\in I}$ as online sub $\alpha$-adjustment, if $\alpha_i^I$ is measurable regarding $\mathcal{F}_{i}$ for all $i\in I$. With this, each $\boldsymbol{\alpha}_I$ defines an online intersection test $\phi_I$ by 
\begin{align}\phi_I =\mathbbm{1}\{\exists i\in I: p_i \leq \alpha_i^I \}.\label{eq:sub_adj}\end{align}
In what follows, we introduce the predictability condition for the family of online sub $\alpha$-adjustments $(\boldsymbol{\alpha}_I)_{I\subseteq \mathbb{N}}$.

\begin{definition}[Predictable family of online sub $\alpha$-adjustments\label{def:pred_fam_alpha_adj}]
A family of online sub $\alpha$-adjustments  $(\boldsymbol{\alpha}_I)_{I \subseteq \mathbb{N}}$ is called \textit{predictable}, if for all $I\subseteq \{1,...,i\}$ and $K=I\cup J$ with $J\subseteq \{k\in \mathbb{N}:k>i\}$ it holds that $\alpha_j^I=\alpha_j^K$ for all $j\in I$.
\end{definition}

 Note that this condition implies that the corresponding family of online intersection tests $\boldsymbol{\phi}=(\phi_I)_{I\subseteq \mathbb{N}}$ defined by \eqref{eq:sub_adj} is predictable as well (Definition \ref{def: predictable family}) and hence the resulting closed procedure $\boldsymbol{d}^{\boldsymbol{\phi}}$ is an online procedure (Theorem \ref{theo:online closure principle}). However, it is not necessary for predictability of $\boldsymbol{\phi}$, as one could also choose $\alpha_j^I<\alpha_j^K$, where $I$ and $K$ are defined as in Definition \ref{def:pred_fam_alpha_adj}. We use Definition \ref{def:pred_fam_alpha_adj}, since we think it is the usual way of constructing predictable intersection tests based on $\alpha$-adjustments, as we will also illustrate in Sections \ref{sec:closed alpha-spending}- \ref{sec:closed addis-spending}. Furthermore, in case of $\alpha_j^I<\alpha_j^K$, the significance levels are not monotone \citep{HBW, romano2005exact}, which is often assumed to obtain consonant intersection tests. Excluding this case helps to define the short-cut in Theorem \ref{theo:consonance procedure} as an $\alpha$-adjustment procedure, which we will show in the remainder of this section.

\begin{remark}
Many online $\alpha$-adjustments $(\alpha_i)_{i\in \mathbb{N}}$ in current literature can be defined by a fixed algorithm $A$ that takes a finite vector of $p$-values as input and outputs an individual significance level such that $\alpha_i=A(p_1,\ldots,p_{i-1})$ for all $i\in \mathbb{N}$. This ensures that $\alpha_i$ is measurable with respect to $\sigma(p_1,\ldots,p_{i-1})\subseteq \mathcal{F}_i$. Note that Definition \ref{def:pred_fam_alpha_adj} is always fulfilled if the same algorithm $A$ is used for each online sub $\alpha$-adjustment procedure, meaning that $\alpha_i^I=A\left((p_j)_{j\in I, j<i}\right)$ for all $i\in I \subseteq \mathbb{N}$. To see this, let $I\subseteq \{1,...,i\}$ and $K=I\cup J$ with $J\subseteq \{j\in \mathbb{N}:j>i\}$. Then, we have $\alpha_l^I=A\left((p_k)_{k\in I, k<l}\right)=A\left((p_k)_{k\in K, k<l}\right)=\alpha_l^K$ for all $l\in I$. This is also what Tian \& Ramdas (2021) \citep{TR} formulated as initial attempt for extending the closure principle to the online setting. However, they did not show that this indeed leads to an online procedure. Furthermore, it is only a special case of our more general online closure principle, as we allow to use different algorithms for each intersection hypothesis and consider general online sub $\alpha$-adjustments or even general online intersection tests. 
\end{remark}

If the family of online intersection tests $(\phi_I)_{I\subseteq \mathbb{N}}$ defined by \eqref{eq:sub_adj} additionally satisfies the consonance property \eqref{eq:consonance}, the short-cut described in Theorem \ref{theo:consonance procedure} can be expressed as an online $\alpha$-adjustment procedure (see Appendix for the proof).

\begin{theorem}\label{corol:consonance procedure}
Assume $(\boldsymbol{\alpha}_I)_{I \subseteq \mathbb{N}}$, where $\boldsymbol{\alpha}_I=(\alpha_i^{I})_{i\in I}$, is a predictable family of online sub $\alpha$-adjustments such that  $\boldsymbol{\phi}=(\phi_I)_{I\subseteq N}$ defined by \eqref{eq:sub_adj} is a family of $\alpha$-level intersection tests with the consonance property. Let us recursively define $I_1=\{1\}$ and $I_i=\{j\in \mathbb{N}: j<i, p_j>\alpha_j^{I_j}\} \cup \{i\}$ for all $i\geq 2$. Then the following three procedures lead to the same decisions:
\begin{enumerate}
\item The online closed procedure $\boldsymbol{d}^{\boldsymbol{\phi}}$.
\item The short-cut $\boldsymbol{d}^{\boldsymbol{\phi},s}=(d^{\boldsymbol{\phi},s}_i)_{i\in \mathbb{N}}$, where $d^{\boldsymbol{\phi},s}_i=\phi_{I_i}$ for all $i\in \mathbb{N}$.
\item The online $\alpha$-adjustment procedure $\boldsymbol{d}=(d_i)_{i\in \mathbb{N}}$, where $d_i=\mathbbm{1}\{p_i\leq \alpha_i^{I_i}\}$ for all $i\in \mathbb{N}$.
\end{enumerate}
\end{theorem}

In the remaining of this section, we apply Theorem \ref{corol:consonance procedure} to construct new online $\alpha$-adjustment procedures based on existing ones. We start with the Alpha-Spending \cite{FS}, deriving a simple improvement of it in Subsection \ref{sec:closed alpha-spending} which serves to exemplify how to use the proposed short-cuts. In Subsection \ref{sec:online-graph}, we show how an online version of the graphical procedure by Bretz et al. \cite{BWBP} can be obtained by the new online closure principle. Finally, we derive an improvement of the ADDIS-Spending under local dependence \cite{TR} (Subsection \ref{sec:closed addis-spending}). 

\subsection{Closed Alpha-Spending\label{sec:closed alpha-spending}}

The Alpha-Spending is an online version of the weighted Bonferroni, meaning that the overall significance level $\alpha$ is split between the individual hypotheses $(H_i)_{i\in \mathbb{N}}$ according to some weights.
\begin{definition}[Alpha-Spending\label{def:alpha-spending} \cite{FS}]
Let $(\gamma_i)_{i\in \mathbb{N}}$ be a non-negative sequence of real numbers with $\sum_{i=1}^{\infty} \gamma_i \leq 1$. For a given $\text{FWER}$ level $\alpha$, \textit{Alpha-Spending} tests for every $i\in \mathbb{N}$ the hypothesis $H_i$ at the individual level $$\alpha_i= \alpha \gamma_i.$$ \end{definition}

The strong FWER control of the Alpha-Spending follows by the Bonferroni inequality \cite{FS}. However, as the Bonferroni, the Alpha-Spending is generally a conservative procedure, meaning that uniform improvements exist that could possibly be obtained by the closure principle. In order to derive a closure of the Alpha-Spending, we first have to formulate an online intersection test based on the Alpha-Spending. Here, we just apply the Alpha-Spending on a subsequence by ignoring the $p$-values that are not contained in it.

\begin{definition}[Alpha-Spending intersection test\label{def:alpha-spending-int}]
Let $(\gamma_i)_{i\in \mathbb{N}}$ be as in Alpha-Spending. The \textit{Alpha-Spending intersection test} $\phi_I$ is defined by \eqref{eq:sub_adj}, where 
$ \alpha_i^{I} = \alpha \gamma_{t_I(i)}$ with  $t_I(i)=|\{j\in I: j\leq i\}|$, for all $I\subseteq \mathbb{N}$.
\end{definition}

We assume that the same $(\gamma_i)_{i\in \mathbb{N}}$ is used for all intersection tests $\phi_I$. Note that for determining $\alpha_i^I$ with $i\in I \subseteq \mathbb{N}$ it is only important how many indices $j<i$ are included in $I$, but we do not need information about the indices that are greater than $i$. This ensures the predictability (Definition \ref{def:pred_fam_alpha_adj}) of  $(\boldsymbol{\alpha}_I)_{I\subseteq \mathbb{N}}$, where $\boldsymbol{\alpha}_I=(\alpha_i^I)_{i\in I}$. However, in general, $(\phi_I)_{I\subseteq \mathbb{N}}$ does not have the consonance property and thus we cannot apply Theorem \ref{corol:consonance procedure}. For example, consider $(\gamma_i)_{i\in \mathbb{N}}=(0,1,0,0,\ldots)$. If $p_2\leq \alpha$, we have $\phi_{\{1,2\}}=1$ but $\phi_{\{1\}}=0$ and $\phi_{\{2\}}=0$. Hence, the consonance property is not satisfied. Also note that for the \enquote{consonantized} intersection tests (see Section \ref{sec:admissibility}), we have $\tilde{\phi}_I=\max\{d_i^{\phi}:i\in I\}=0$ for all $I\subseteq \mathbb{N}$, since $\phi_{\{i\}}=0$ for all $i\in \mathbb{N}$. This exemplifies how requiring consonance can help to identify the inefficiency of closed procedures \citep{romano2011consonance}. We can ensure to obtain consonant Alpha-Spending intersection tests by choosing $(\gamma_i)_{i\in \mathbb{N}}$ to be non-increasing. To see this, consider $I\subseteq \mathbb{N}$ with $\phi_I=1$. Then there exists an $i\in I$ such that $p_i \leq \alpha_i^I=\alpha \gamma_{t_I(i)}$. Now for $J\subseteq I$ with $i\in J$, it holds that $t_J(i)\leq t_I(i)$. If $(\gamma_i)_{i\in \mathbb{N}}$ is non-increasing, this implies $p_i\leq  \alpha \gamma_{t_J(i)}=\alpha_i^J$ and hence $\phi_{J}=1$. Note that it is fairly common to choose $(\gamma_i)_{i \in \mathbb{N}}$ to be non-increasing, since it needs to converge to $0$ anyway. This leads to the following new online closed $\alpha$-adjustment procedure.

\begin{definition}[Closed Alpha-Spending\label{def:closed-alpha-spending}] Let $(\gamma_i)_{i \in \mathbb{N}}$ be as in Alpha-Spending but non-increasing. \textit{Closed Alpha-Spending} updates the individual significance levels as follows
$$\alpha_i = \alpha \gamma_{t(i)}, $$
 $\text{where } t(i)= 1 + \sum_{j=1}^{i-1} (1- d_j)$ and $d_j=\mathbbm{1}\{p_j \leq \alpha_j\}$.
\end{definition}

\begin{prop}
    Closed Alpha-Spending controls the FWER in the strong sense.
\end{prop}
\begin{proof}
    Let $(\phi_I)_{I\subseteq \mathbb{N}}$ be a family of Alpha-Spending intersection tests based on the same non-increasing $(\gamma_i)_{i\in \mathbb{N}}$.  Due to Theorem \ref{corol:consonance procedure}, the individual significance levels of the resulting closed procedure are given by $\alpha_i^{I_i}=\alpha \gamma_{t_{I_i}(i)}$, where 
    $$t_{I_i}(i)=1+|\{ j<i: p_j>\alpha_j^{I_j}\}|=1+\sum_{j=1}^{i-1} (1-\mathbbm{1}\{p_j\leq \alpha_j^{I_j}\}).$$ Hence, the FWER control follows by the online closure principle (Theorem \ref{theo:online closure principle}).
\end{proof}
 
It is easy to verify that the closed Alpha-Spending (Definition \ref{def:closed-alpha-spending}) is a uniform improvement of the Alpha-Spending (Definition \ref{def:alpha-spending}). Alternative closures of the Alpha-Spending procedure can be derived, which are often online variants of the Bonferroni-based closed procedures \cite{HBW}.

\begin{remark} By applying the Alpha-Spending intersection test (Definition \ref{def:alpha-spending-int}) to every intersection hypothesis $H_I$ the significance level might not be fully exhausted when $I$ is finite. However, this is inevitable to obtain a predictable family of online intersection tests. Suppose a closed procedure where for each intersection hypothesis $H_I$ an online sub $\alpha$-adjustment 
$(\alpha_i^I)_{i\in I}$ is chosen such that $\sum_{i \in I} \alpha_i^I=\alpha$. Then the intersection hypothesis $H_{\{i\}}$ would be rejected if $p_i \leq \alpha$. Now consider  $J\subseteq \mathbb{N}$ with $\min(J)=i$. The only option such that $\phi_{\{i\}}=1$ implies $\phi_{J}=1$ is to choose $\alpha_i^{J}=\alpha$ and therefore $\alpha_{j}^{J}=0$ for all $j\in J\setminus \{i\}$. The resulting online closed procedure is the online analogue of the fixed sequence procedure \cite{MHL}: reject $H_i$ if $p_1\leq \alpha, \ldots, p_i \leq \alpha$. Thus, if $p_j >\alpha$ for a $j\in \mathbb{N}$, no hypothesis $H_k$ with $k\geq j$ can be rejected. This is unfavourable in most online scenarios. 
\end{remark}


\subsection{Online-Graph\label{sec:online-graph}}

In classical multiple testing, the graphical procedure by Bretz et al. \cite{BWBP} has grown in popularity over the last years due to its easy interpretability and high power. Its FWER control is shown by the closure principle and by applying a weighted Bonferroni test to each intersection hypothesis. Tian and Ramdas \cite{TR} have built on the result and extended the graphical procedure to the online setting which led to an Online-Graph (they termed it Online-Fallback in \cite{TR}). In this subsection, we give the formal description of the aforementioned graphical procedure and afterwards sketch how the Online-Graph can be obtained by the proposed online closure principle. 

\begin{definition}[Graphical procedure \cite{BWBP}\label{def:offgraph}]
Let $\mathcal{H}=\{H_1,\ldots,H_m\}$ be the set of hypotheses to be tested, $\{p_1,\ldots,p_m\}$ the corresponding $p$-values and $\{\alpha_1,\ldots,\alpha_m\}$ the initial allocation of the overall significance level, where for all $i\in \{1,\ldots,m\}$: $\alpha_i=\alpha \gamma_i$ such that $\sum_{i =1}^{m} \gamma_i \leq 1$. In addition, let $\boldsymbol{G}=(g_{i,j})_{i,j\in \{1,\ldots,m\}}$ be a matrix containing non-negative weights such that $g_{i,i}=0$ and $\sum_{j=1}^{m} g_{i,j} \leq 1$ for all $i\in \{1,\ldots,m\}$. Then the graphical $\alpha$-adjustment is defined by the following stepwise algorithm:

\begin{enumerate}
\setcounter{enumi}{-1}
\item Set $I=\{1,\ldots,m\}$.
\item\label{alg: step_1_graph} Let $i=\argmin_{j\in I} \frac{p_j}{\alpha_j}$. If $p_i>\alpha_i$, stop and accept all hypotheses that have not been rejected yet. 
\item Reject $H_i$ and update $I$, the individual significance levels and the weights as follows:
\begin{align*}I\rightarrow I\setminus \{i\}, \text{ } \alpha_j \rightarrow \begin{cases} \alpha_j + \alpha_i g_{i,j}, \text{ } & j \in I  \\ 0,  & \text{otherwise} \end{cases}, \text{ } g_{j,k} \rightarrow \begin{cases} \frac{g_{j,k} + g_{j,i} g_{i,k}}{1-g_{j,i} g_{i,j}}, \text{ } & j,k\in I, \text{ }j\neq k \\ 0, & \text{otherwise} \end{cases}. 
\end{align*}  \label{alg: step_2_graph}
\item If $|I|\geq 1$, go to step $1$. Else, stop.
\end{enumerate}
\end{definition}

\begin{figure}[htbp]
 	\begin{center}
 			\centering
 			\includegraphics[width=19.5cm,height=6.5cm,keepaspectratio]{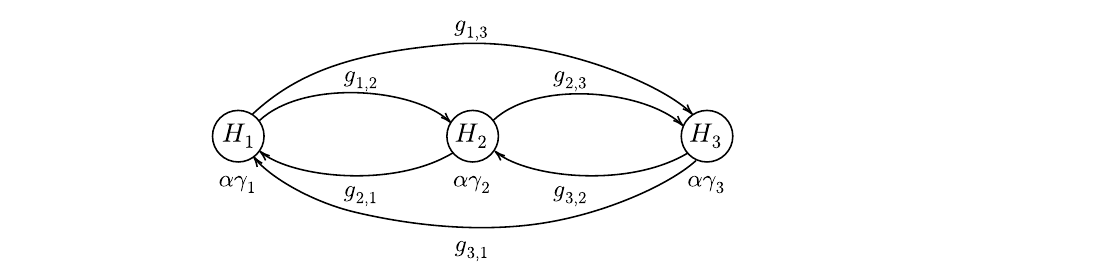}
 	\end{center}
	\caption{Illustration of the graphical procedure by \cite{BWBP} for $m=3$ hypotheses.\label{fig:offline-graph}}
 \end{figure}

 In Figure \ref{fig:offline-graph}, the graphical procedure (Definition \ref{def:offgraph}) is illustrated for $m=3$ hypotheses. Below each hypothesis is the initial individual significance level. The arrows represent the allocation of the weights after rejecting a hypothesis. After each rejection, the graph is updated according to Step \ref{alg: step_2_graph} of the graphical procedure.

 In online multiple testing, only future hypotheses are allowed to benefit from a rejection. Since there are no arrows pointing back, the weights $g_{i,j}$ do not need to be updated at any step. To derive the Online-Graph by the online closure principle, we first define online intersection tests for each $I\subseteq \mathbb{N}$ that are based on Alpha-Spending (Definition \ref{def:alpha-spending}). The idea is to start with the same initial allocation of the significance level as in Alpha-Spending which is given by $(\gamma_i)_{i\in \mathbb{N}}$. That means $H_{\mathbb{N}}$ is rejected if $p_i\leq \alpha_i^{\mathbb{N}}=\alpha \gamma_i$ for at least one $i\in \mathbb{N}$. Now consider $H_I$ for some index set $I\subseteq \mathbb{N}$. If $j\notin I$, then the individual significance level of $H_j$ is distributed to the future hypotheses according to weights $(g_{j,i})_{i=j+1}^{\infty}$ such that $\sum_{i=j+1}^{\infty} g_{j,i}\leq 1$. Since significance level is only assigned to future hypotheses, we have $\alpha_1^I=\alpha \gamma_1$ for all $I\subseteq \mathbb{N}$ with $1\in I$ and the other levels can be defined recursively by
$$\alpha_i^I=\alpha \left(\gamma_i+\sum_{j<i,j\notin I} g_{j,i} \alpha_j^{(I \cup j)} \right) \quad (i\in I). $$
The above derivation implies that $\sum_{i\in I} \alpha_i^I\leq \alpha \sum_{i\in \mathbb{N}} \gamma_i \leq \alpha$, such that each of these online sub $\alpha$-adjustments $\boldsymbol{\alpha}_I=(\alpha_i^I)_{i\in I}$ defines an online $\alpha$-level intersection test $\phi_I$ by \eqref{eq:sub_adj}. In addition, the predictability of $(\boldsymbol{\alpha}_I)_{I\subseteq \mathbb{N}}$ and consonance of $\boldsymbol{\phi}=(\phi_I)_{I\subseteq \mathbb{N}}$ can easily be verified. The Online-Graph is then obtained as the short-cut of the online closed procedure $\boldsymbol{d}^{\boldsymbol{\phi}}$ and thus is defined by 
$$\alpha_i=\alpha_i^{I_i}=\alpha \left(\gamma_i+\sum_{j<i} g_{j,i} \alpha_j d_j \right) \quad (i\in \mathbb{N}),$$
where $d_j=\mathbbm{1}\{p_j\leq \alpha_j\}$ (see Theorem \ref{corol:consonance procedure}). 

\begin{figure}[htbp]
	\begin{center}
			\centering
			\includegraphics[width=19.5cm,height=6.5cm,keepaspectratio]{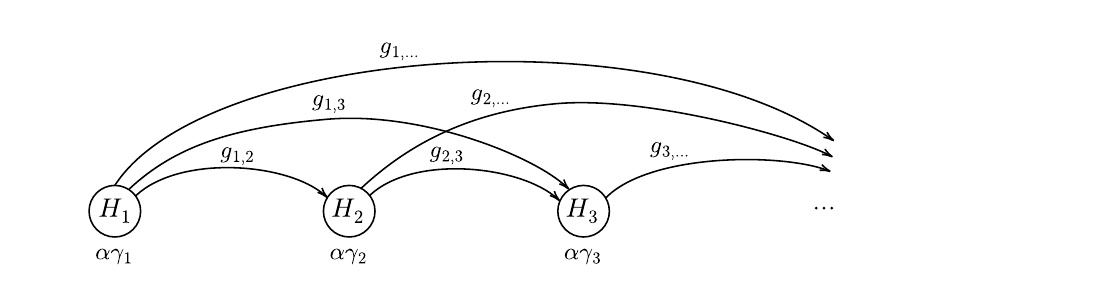}
	\end{center}
	\caption{Illustration of the Online-Graph.\label{fig:online-graph}}
\end{figure}

Figure \ref{fig:online-graph} shows the illustration of the Online-Graph. Compared to Figure \ref{fig:offline-graph}, the arrows only point to future hypotheses. As already mentioned, this means that only future significance levels are updated after a rejection. In addition, the weights $(g_{j,i})_{i=j+1}^{\infty}$, $j\in \mathbb{N}$, remain the same over the entire testing process and do not have to be updated. The points at the end indicate that there is an infinite number of future hypotheses.

\begin{remark}\hphantom{1}
\begin{enumerate}
    \item Note that not all graphs with an infinite number of hypotheses are special cases of this Online-Graph. One could also think of graphical procedures that allocate significance levels to previous hypotheses. For example, suppose that after each rejection the significance level of the rejected hypothesis is distributed to the first hypothesis. This could be the case, for instance, if the hypothesis $H_1$ is of main interest and the rejection of one of the future hypotheses should increase the probability of rejecting $H_1$. Obviously, this does not define an online procedure. Nevertheless, it can be written as a closed procedure resulting from the following Alpha-Spending based online intersection test:
$\phi_I=1$, if $p_1\leq \alpha\left(\gamma_1 + \sum_{i\notin I} \gamma_i \right) \mathbbm{1}\{1\in I\}$ or $p_i\leq \alpha \gamma_i$ for at least one $i\in I$. 
This again clarifies that the predictability of $(\phi_I)_{I\subseteq \mathbb{N}}$ (Definition \ref{def: predictable family}) is needed in order to obtain online closed procedures.
\item There is a strong connection between the Online-Graph and closed Alpha-Spending (Definition \ref{def:closed-alpha-spending}). If  we would choose $\tilde{g}_{j,i}=(\gamma_{t(j)+i-j-1}-\gamma_{t(i)+i-j})/\gamma_{t(j)}$ as weights for the Online-Graph, where $t(j)=1+\sum_{k=1}^{j-1}(1-d_k)$, both procedures would be the same. However, in general the weights are not allowed to depend on the previous rejections and thus we would not consider closed Alpha-Spending as a special case of the Online-Graph. Nevertheless, in some cases both procedures collapse. For example, if $\gamma_i=q^i(1-q)/q$ for some $q\in (0,1)$, we have $\tilde{g}_{j,i}=q^{i-j} (1-q)/q =\gamma_{i-j}$, which is independent of the data.
\end{enumerate}
\end{remark}


\subsection{Closed ADDIS-Spending\label{sec:closed addis-spending}}

Although the Online-Graph is a uniform improvement of the Alpha-Spending and its offline version is one of the most popular procedures in classical multiple testing, in particular for clinical trials, Tian and Ramdas (2021) \cite{TR} claimed that their proposed ADDIS-Spending is to be preferred in the online setting. 

ADDIS-Spending combines the multiple testing approaches of discarding large $p$-values using a parameter $\tau$ \citep{ZSS} and adapting to the number of false hypotheses using a parameter $\lambda$ \citep{schweder1982plots}. Suppose that the $p$-values are uniformly valid, which means $\mathbb{P}(p_i\leq x\tau|p_i\leq \tau) \leq x $ for all $x,\tau\in [0,1]$, $\mathbb{P}\in H_i$, and let $\tau_i\in (0,1]$, $\lambda_i\in [0,\tau_i)$ and $\alpha_i\in [0,\tau_i)$ be fixed parameters for all $i\in \mathbb{N}$. Then, by the Bonferroni inequality, for all $\mathbb{P}\in \mathcal{P}$,  we have:
\begin{align*}\text{FWER}_{\mathbb{P}}&\leq \sum_{i\in I_0^{\mathbb{P}}}\mathbb{P}(p_i\leq \alpha_i) = \sum_{i\in I_0^{\mathbb{P}}}\mathbb{P}(p_i\leq \alpha_i | p_i\leq \tau_i) \mathbb{P}(p_i\leq \tau_i) \leq \sum_{i\in I_0^{\mathbb{P}}}\frac{\alpha_i}{\tau_i} \mathbb{P}(p_i\leq \tau_i) \nonumber \\ &\leq \sum_{i\in I_0^{\mathbb{P}}}\frac{\alpha_i}{\tau_i} \mathbb{P}(p_i\leq \tau_i) \frac{\mathbb{P}(p_i> \lambda_i|p_i\leq \tau_i)}{1-\lambda_i/\tau_i}\leq\mathbb{E}_{\mathbb{P}} \left[ 
 \sum_{i\in \mathbb{N}} \frac{\alpha_i}{\tau_i-\lambda_i} \mathbbm{1}\{\lambda_i < p_i \leq \tau_i\}\right],\end{align*}
 where the second and third inequality follow from the uniform validity of the $p$-values. We restricted to fixed parameters for simplicity. However, the same calculation can be done when $\tau_i$, $\lambda_i$ and $\alpha_i$ only depend on information that is independent of $p_i$, by conditioning on this information \citep{TR}. Also note that in the case of $\tau_i=1$, the uniform validity would not be needed and thus this assumption is only required for the discarding and not the adaptive part \citep{TR}. Anyway, uniform validity is fulfilled in many settings \citep{ZSS,TR}. The above calculation shows that in order to control the FWER, it is sufficient to ensure
  \begin{align}
      \sum_{i\in \mathbb{N}} \frac{\alpha_i}{\tau_i-\lambda_i} \mathbbm{1}\{\lambda_i < p_i \leq \tau_i\} \leq \alpha. \label{eq:condition_ADDIS}
  \end{align}
  The idea is that we can reuse the significance level if $P_i\leq \lambda_i$ or $P_i> \tau_i$, but need to subtract the larger significance level $\alpha_i/(\tau_i-\lambda_i)$ if $\lambda_i<P_i\leq \tau_i$. Since $p$-values corresponding to true hypotheses tend to be large and $p$-values corresponding to false hypotheses tend to be small, we expect this tradeoff to be useful. In order to exploit this condition, the individual significance levels need to depend on information about the $p$-values observed so far. Since $\tau_i$, $\lambda_i$ and $\alpha_i$ need to be independent of $p_i$, more assumptions on the dependence structure of the $p$-values are needed. 
 
 An example is local dependence \cite{ZRJ} which allows $p$-values that are close together in time to depend on each other while $p$-values that are further apart are independent. This is an intuitive condition, because one would think that $p$-values that are closer in time are stronger related than those with a large time gap. For example, local dependence encompasses batch dependence, where $p$-values within one batch may depend on each other but $p$-values from different batches are independent. This is the case in practice, for instance, if the used data is replaced by independent data after a period of time. Mathematically, local dependence is defined as follows.

\begin{definition}[Local dependence \cite{ZRJ}]
Let $(l_i)_{i\in \mathbb{N}}$ be a fixed sequence of parameters such that $l_i\in \{0,1,\ldots,i-1\}$ and $l_{i+1} \leq l_i +1$ for all $i\in \mathbb{N}$. A sequence of $p$-values $(p_i)_{i\in \mathbb{N}}$ is called \textit{locally dependent} with the lags $(l_i)_{i\in \mathbb{N}}$, if $\forall i \in \mathbb{N}$ holds:
$$ p_i \perp p_{i-l_i-1}, p_{i-l_i-2}, \ldots, p_1.$$
\end{definition}
If $l_i=0$ for all $i\in \mathbb{N}$, all $p$-values are independent. In the other extreme case, $l_i=i-1$ for all $i\in \mathbb{N}$, all $p$-values are dependent. Although it is assumed that the lags are constant parameters, in practice, one does not have to know all $l_i$ at the beginning of the evaluation. However, one must determine $l_i$ before testing hypothesis $H_i$ without using the data itself. Tian and Ramdas (2021) \cite{TR} proposed the following online procedure that satisfies condition \eqref{eq:condition_ADDIS} under local dependence and thus controls the FWER in the strong sense when the $p$-values are uniformly valid.
\begin{definition}[ADDIS-Spending under local dependence\label{def_addis_spa_lok}]
Assume local dependence with lags $(l_i)_{i\in \mathbb{N}}$ and let $(\gamma_i)_{i\in \mathbb{N}}$ be a non-increasing sequence of weights for an Alpha-Spending (Definition \ref{def:alpha-spending}). In addition, let $(\tau_i)_{i\in \mathbb{N}}$ and $(\lambda_i)_{i\in \mathbb{N}}$ be sequences of random variables such that $\tau_i\in (0,1]$ and $\lambda_i\in [0,\tau_i)$ are measurable with respect to $\mathcal{G}_{i-l_i-1}$ for all $i\in \mathbb{N}$, where $\mathcal{G}_{i-l_i-1}=\sigma(\{p_1,\ldots,p_{i-l_i-1}\})$. The \textit{ADDIS-Spending} under local dependence updates the individual significance levels as follows
\begin{align*}\alpha_i = \alpha (\tau_i-\lambda_i) \gamma_{t(i)}, 
\end{align*} $\text{where } t(i)=1+l_i+\sum_{j=1}^{i-l_i-1} (s_j-c_j)$, $s_j=\mathbbm{1}\{p_j \leq \tau_j\}$ and $c_j=\mathbbm{1}\{p_j \leq \lambda_j\}$.
\end{definition}
  
 We investigate next, whether the above ADDIS-Spending procedure can be improved by an online closed procedure. For this closed procedure, we first define an ADDIS-Spending intersection test. To this end, we define the index set $L_i\coloneqq \{i-l_i,\ldots,i-1\}$ of previous $p$-values that depend on $p_i$. 

\begin{definition}[ADDIS-Spending intersection test\label{def: global ex-addis-spa}]
Assume local dependence with lags $(l_i)_{i\in \mathbb{N}}$ and let $(\gamma_i)_{i\in \mathbb{N}}$, $(\tau_i)_{i\in \mathbb{N}}$ and $(\lambda_i)_{i\in \mathbb{N}}$ be as in ADDIS-Spending (Definition \ref{def_addis_spa_lok}). The \textit{ADDIS-Spending intersection test} $\phi_I$ is defined by \eqref{eq:sub_adj},  where
$$\alpha_i^I=\alpha (\tau_i-\lambda_i) \gamma_{t_I(i)}$$
with $t_I(i)=1+ |L_i\cap I|+\sum_{j\leq i-l_i-1, j\in I} (s_j-c_j)$ for all $i\in I$.
\end{definition}

FWER control of the ADDIS-Spending directly implies that the ADDIS-Spending intersection test is an online $\alpha$-level intersection test under local dependence and uniform validity of $p$-values. In addition, as with the Alpha-Spending, it can easily be verified that the family of online sub $\alpha$-adjustments $(\boldsymbol{\alpha}_I)_{I\subseteq \mathbb{N}}$, where $\boldsymbol{\alpha}_I=(\alpha_i^I)_{i\in I}$, is predictable and the corresponding family of ADDIS-Spending intersection tests $\boldsymbol{\phi}=(\phi_I)_{I\subseteq \mathbb{N}}$ satisfies the consonance property when the same parameters $(\gamma_i)_{i\in \mathbb{N}}$, $(\tau_i)_{i\in \mathbb{N}}$ and $(\lambda_i)_{i\in \mathbb{N}}$ are used for each intersection test. With that, the short-cut of the closed procedure can be obtained by Theorem \ref{corol:consonance procedure}. 

\begin{definition}[Closed ADDIS-Spending\label{def:closed_addis_spending}]
Assume local dependence with lags $(l_i)_{i\in \mathbb{N}}$ and let $(\gamma_i)_{i\in \mathbb{N}}$, $(\tau_i)_{i\in \mathbb{N}}$ and $(\lambda_i)_{i\in \mathbb{N}}$ be as in ADDIS-Spending (Definition \ref{def_addis_spa_lok}). \textit{Closed ADDIS-Spending} updates the individual significance levels as follows 
\begin{align*}\alpha_i = \alpha (\tau_i-\lambda_i) \gamma_{t(i)}, 
\end{align*} $\text{where } t(i)=1+\sum_{j=1}^{i-l_i-1} (s_j-\max\{c_j,d_j\})+\sum_{j=i-l_i}^{i-1} (1-d_j)$ with $d_j=\mathbbm{1}\{p_j\leq \alpha_j\}$.
\end{definition}

\begin{prop}
    Closed ADDIS-Spending controls the FWER in the strong sense under local dependence when the $p$-values are uniformly valid.
\end{prop}
\begin{proof}
    Let $(\phi_I)_{I\subseteq \mathbb{N}}$ be a family of ADDIS-Spending intersection tests based on the same parameters $(\gamma_i)_{i\in \mathbb{N}}$, $(\tau_i)_{i\in \mathbb{N}}$ and $(\lambda_i)_{i\in \mathbb{N}}$.  Due to Theorem \ref{corol:consonance procedure} the individual significance levels of the resulting closed procedure are given by $\alpha_i^{I_i}=\alpha (\tau_i-\lambda_i) \gamma_{t_{I_i}(i)}$, where 
    \begin{align*}t_{I_i}(i)&=1+\Big\vert L_i\cap \left\{j<i:\alpha_j^{I_j}>p_j\right\}\Big\vert + \sum_{j\leq i-l_i-1, p_j>\alpha_j^{I_j}} (s_j-c_j) \\ &= 1+ \sum_{j=i-l_i}^{i-1} (1-\mathbbm{1}\{p_j\leq \alpha_j^{I_j}\}) + \sum_{j=1}^{i-l_i-1} (s_j-c_j)(1-\mathbbm{1}\{p_j\leq \alpha_j^{I_j}\}).\end{align*} Hence, the FWER control follows by the online closure principle (Theorem \ref{theo:online closure principle}). 
\end{proof}
Note that $\sum_{j=i-l_i}^{i-1} (1-d_j)\leq l_i$ and therefore $t(i)$ in Definition \ref{def:closed_addis_spending} is never larger than $t(i)$ in Definition \ref{def_addis_spa_lok}. Since  $(\gamma_i)_{i\in \mathbb{N}}$ is non-increasing, closed ADDIS-Spending never rejects  less hypotheses than ADDIS-Spending. Furthermore, if there are dependent $p$-values, which means $l_i>0$ for some $i\in \mathbb{N}$, closed ADDIS-Spending is a real uniform improvement of ADDIS-Spending. In this case, ADDIS-Spending can only gain significance level from independent $p$-values, while closed ADDIS-Spending additionally allows to gain from rejections of dependent hypotheses. For example, let $p_1$ and $p_2$ depend on each other and $p_1 \leq \alpha_1$. Then $H_2$ is tested at level $\alpha (\tau_2-\lambda_2) \gamma_2$ using ADDIS-Spending and at level $\alpha (\tau_2-\lambda_2) \gamma_1$ using closed ADDIS-Spending. If $\lambda_i < \alpha_i$ for some $i\in \mathbb{N}$, we have an additional improvement, since $\max\{c_i,d_i\}>c_i$ if $\lambda_i<P_i\leq \alpha_i$. In particular, for $\lambda_i=0$ closed ADDIS-Spending provides a uniform improvement of Discard-Spending \citep{TR} under independent $p$-values. 



\section{Simulation study\label{sec:sim}}

In this section, we aim to quantify the gain in power when using closed ADDIS-Spending instead of ADDIS-Spending and show its FWER control by means of simulations. For this purpose, we first describe the simulation design and then show the results. We considered similar simulation scenarios as in \cite{TR}, but generated locally dependent $p$-values instead of independent $p$-values.

\subsection{Simulation design\label{subsimsetup}}

We simulated trials where $n=1000$ null hypotheses
$(H_i)_{i\in \{1,\ldots,n\}}$ are tested sequentially. We assume that the local dependence structure of the $p$-values  is given by finite batches $B_i$, $i\in \mathbb{N}$, with a fixed batch-size $b\in \mathbb{N}$. That means we have batches $B_{1}=\{p_1,\ldots,p_b\}$, $B_2=\{p_{b+1},\ldots,p_{2b}\}$ and so forth, and the $p$-values within one batch depend on each other while $p$-values from different batches are independent. For this simulation, we considered $b\in \{1,10,25,100\}$. 

Let $X^{1:b},X^{(b+1):2b},\ldots,X^{(n-b+1):n} \stackrel{i.i.d.}{\sim} N_b(\mu_0,\Sigma)$, where $X^{j:i}=(X_j,\ldots,X_i)^T$ for $i\geq j$, and where $N_b$ is the $b$-dimensional normal distribution, $\mu_0= (0,\ldots,0)^T\in \mathbb{R}^{b}$ and $\Sigma=(\sigma_{ij})_{i,j=1,\ldots,b}\in \mathbb{R}^{b\times b}$ with $\sigma_{ii}=1$ and $\sigma_{ij}=\rho\in (0,1)$ for all $i\in \{1,\ldots,b\}$ and $j\neq i$.
 For each $i\in \{1,\ldots,n\}$, we test the null hypothesis $H_i: \mathbb{E}[Z_i] \leq 0$,  where $Z_i=X_i+\mu_A$, $\mu_A>0$, with probability $\pi_A\in (0,1)$ and $Z_i=X_i+\mu_N$, $\mu_N\leq 0$, otherwise.
 The $p$-values are calculated by $p_i=\Phi(-Z_i)$, where $\Phi$ is the cumulative distribution function (CDF) of a standard normal distribution. Thus, for a $p$-value of a true hypothesis $p_i=\Phi(-X_i-\mu_N)$, $i\in I_0^{\mathbb{P}}$, and $x\in [0,1]$: \begin{align*} \mathbb{P}(p_i \leq x)=\mathbb{P}(X_i {\leq \Phi^{-1}(x)+\mu_N)}=\Phi(\Phi^{-1}(x)+\mu_N).\end{align*} If $\mu_N=0$, we have uniformly distributed null $p$-values which means that $\mathbb{P}(p_i \leq x)=x$ for all $i\in I_0^{\mathbb{P}}$ and $x\in [0,1]$ while a null $p$-value $p_i$, $i\in I_0^{\mathbb{P}}$, is said to be conservative, if $\mathbb{P}(p_i \leq x)<x$ for some $x\in [0,1]$ \cite{ZSS}. Since $\Phi(x)$ is increasing in $x$, the null $p$-values are conservative if and only if $\mu_N<0$ and the conservativeness grows with decreasing $\mu_N$. The parameters $\pi_A$ and $\mu_A$ can be interpreted as proportion of false hypotheses and strength of the alternative, respectively.

For each considered scenario, we simulated $2000$ independent trials and estimated the power and FWER using  closed ADDIS-Spending and ADDIS-Spending.

\subsection{Comparison of closed ADDIS-Spending and ADDIS-Spending through simulations\label{sec:sim_results}} 

We compared the results when using closed ADDIS-Spending and ADDIS-Spending with respect to FWER and power for different batch-sizes and proportions of false null hypotheses. The FWER is represented by the lines below the global significance level $\alpha= 0.2$ and the power by the lines above it. Thereby, the power is defined as the expected number of  rejected hypotheses among all false hypotheses. Both procedures are applied with the parameters $\gamma_i = \frac{6}{\pi^2 i^2}$, $\lambda_i=0.3$ and $\tau_i=0.8$ for all $i\in \mathbb{N}$. We do not claim that this choice leads to the highest possible power, but it works well to show the differences between the presented procedures. 

\begin{figure}
\centering
\includegraphics[width=12.5cm]{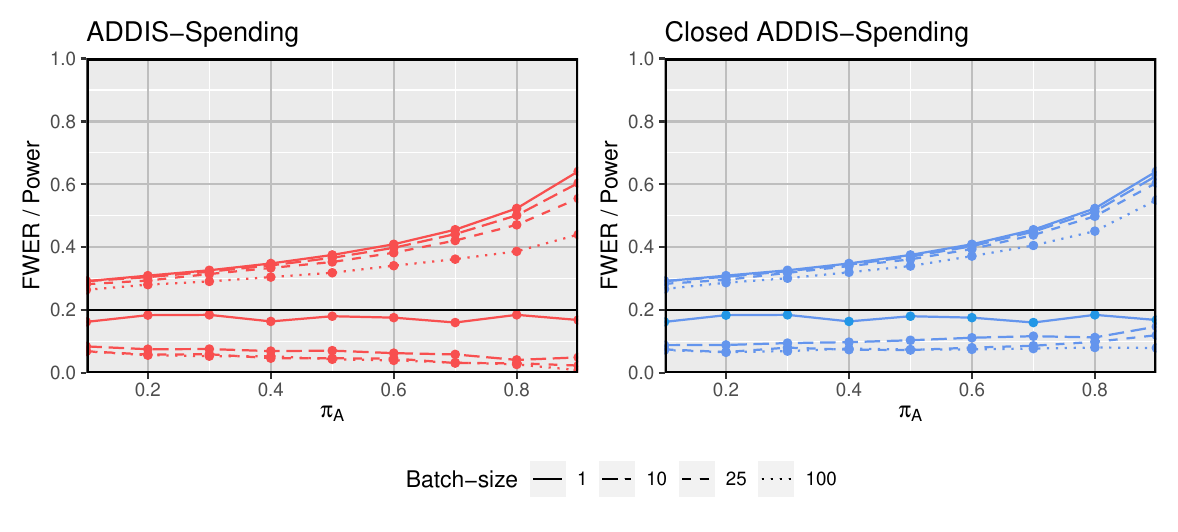}%
\caption{Comparison of ADDIS-Spending and closed ADDIS-Spending in terms of power and FWER ($\alpha=0.2$) based on locally dependent $p$-values for different batch-sizes and proportions of false null hypotheses ($\pi_A$); $n=1000$, $\mu_N=0$, $\mu_A=4$ and $\rho=0.8$ in both plots.\label{fig:simulation_exact}}\end{figure}

The results with uniformly distributed null $p$-values and $\mu_A=4$ can be found in Figure \ref{fig:simulation_exact}. As discussed before, closed ADDIS-Spending and ADDIS-Spending coincide under independence of the $p$-values ($b=1$), by this, the solid lines are identical. However, when some of the $p$-values become dependent ($b>1$), the power and FWER decrease drastically using ADDIS-Spending, while closed ADDIS-Spending decelerates this decrease such that a higher power is obtained. In addition, we see that closed ADDIS-Spending exhausts the FWER more. Simulations for other parameter choices of $n$, $\mu_N$ and $\mu_A$ can be found in the Appendix. In all cases, no differences in the behaviour of closed ADDIS-Spending and ADDIS-Spending are observed compared to those shown in  Figure \ref{fig:simulation_exact}.

\section{Application on real data}
In this section, we apply the presented procedures to real data. First, we consider the IMPC dataset, which aims to identify the influence on the phenotype of each protein-coding mouse gene \citep{mouse} and which is a standard dataset used in the online multiple testing literature \cite{robertson2022online, TR}. Second, we apply the procedures on the RECOVERY platform trial \citep{sandercock2022experiences}. Here, several treatments for severe COVID-19 diseases are tested against a standard of care. 
The two datasets have significant differences. The IMPC data includes several thousand experiments, while the RECOVERY trial has only tested 12 treatments to date. It is important to note that we are not attempting to draw any conclusions from these datasets; they are merely being used for illustrative purposes. 

\subsection{IMPC data\label{sec:IMPC}}
The International Mouse Phenotyping Consortium (IMPC) is coordinating a large-scale study to determine the function of all protein-coding mouse genes.  To this end, each gene is systematically knocked out and the effect on the phenotype is explored. As the dataset grows over time due to the testing of additional genes, the use of online multiple testing procedures is appropriate \citep{robertson2022online, TR}.  Our evaluation is based on the dataset in the Zenodo repository \url{https://zenodo.org/record/2396572} 
 \citep{robertson2019onlinefdr}. The contained $p$-values resulted from the analysis in \cite{karp2017prevalence} and follow a batch dependence structure since the same group of mice was used for testing several consecutive hypotheses \citep{TR}. In our evaluation, we restrict to $5000$ of these $p$-values that are arranged in $84$ batches. 

We compare the number of rejections obtained by Alpha-Spending, closed Alpha-Spending, Online-Graph, ADDIS-Spending and closed ADDIS-Spending at the FWER level $\alpha=0.2$. We choose $(\gamma_i)_{i\in \mathbb{N}}$ such that $\gamma_i\propto 1/i^{h}$ for all $i\in \mathbb{N}$ and $\sum_{i\in \mathbb{N}} \gamma_i=1$, where $h\in \{1.05,1.1,1.3\}$. Note that the larger the $h$, the faster $(\gamma_i)_{i\in \mathbb{N}}$ converges to $0$. As done in \citep{TR}, we set $\tau_i=0.8$ and $\lambda_i=0.16$ for the ADDIS procedures. In addition, we choose $g_{j,i}=\gamma_{i-j}$ for the Online-Graph.

The results are summarized in Table \ref{table:results_impc}. As expected, the Alpha-Spending leads to the least rejections. Closed Alpha-Spending and Online-Graph performed similarly. However, Online-Graph led to more rejections when $(\gamma_i)_{i\in\mathbb{N}}$ decreased faster ($h=1.3$), while closed Alpha-Spending was superior in case of a slowly decreasing $(\gamma_i)_{i\in\mathbb{N}}$ ($h=1.05$). Both procedures were outperformed by ADDIS-Spending, which was further improved by closed ADDIS-Spending.

\begin{table}[h!]
\centering
\caption{Number of rejections obtained by different procedures applied on IMPC data. }
\begin{tabular}[h]{|c|c|c|c|}
\hline
\multirow{2}{*}{\textbf{Procedure}}  & \multicolumn{3}{c|}{\textbf{Number of rejections}} \\
  & $h=1.3$ & $h=1.1$ & $h=1.05$  \\
  \hline
   Alpha-Spending & $1348$ & $1404$ & $1404$  \\
  Closed Alpha-Spending & $1394$ & $1426$ & $1427$ \\ 
  Online-Graph & $1424$ & $1426$ & $1423$  \\ 
  ADDIS-Spending & $1427$ & $1459$ & $1454$  \\
  Closed ADDIS-Spending & $1434$ & $1465$ & $1460$ \\
  \hline
\end{tabular}
\label{table:results_impc}
\end{table}

\subsection{RECOVERY platform trial\label{sec:RECOVERY}}
In a platform trial, several treatment arms $T_1, T_2,\ldots$ are compared to the same control group. In contrast to multi-arm trials, the treatment arms do not enter or leave the trial at the same time and the total number of treatments under evaluation is not pre-defined, leading to an online testing problem \citep{Retal}. Usually, concurrent control data is used, meaning that a treatment arm is only compared to control patients that were randomised while the treatment arm was in the platform. This leads to a local dependence
structure of the $p$-values, since treatment arms that overlap share some control data for testing and those that do
not overlap can be considered as independent.

In this section, we compare the rejections achieved by the considered methods when applied to a real ongoing platform trial. The Randomised Evaluation of COVID-19 Therapy (RECOVERY) trial has already tested twelve treatments for severe COVID-19 diseases against a standard of care, while another one is currently recruiting \cite{sandercock2022experiences}. 
All $p$-values are available at the website \url{https://www.recoverytrial.net/}. The overlapping structure is illustrated in a publication by the data monitoring committee \cite{sandercock2022experiences}. 

We apply the same procedures as in Section \ref{sec:IMPC}. However, since the required evidence in such clinical trials is usually higher, we set $\alpha=0.05$. In addition, we choose $\gamma_i =i^q (1-q)/q$, $i\in \mathbb{N}$, for $q\in \{0.6,0.7,0.8\}$. 
This change is because harmonic sequences tend to decrease very fast at the beginning of the sequence. This is negligible if the $h$ is rather low, as in Section \ref{sec:IMPC}. However, in this example, we would choose larger $h$ as the number of hypotheses is much lower. This is why we expect a geometric sequence to perform better in low-scale settings. Note that in this case, $(\gamma_i)_{i\in \mathbb{N}}$ decreases faster for lower $q$.

We compare the number of rejections and the current individual significance level $\alpha_{13}$ that would be used to test the next treatment, which is already in the trial but has not yet finished recruitment  (see Table \ref{table:results_recovery}). The behavior of the procedures looks similar as in Section \ref{sec:IMPC}. However, while the number of rejections does not differ much, closed ADDIS-Spending tests the current hypothesis at the highest level such that the differences between the number of rejections will possibly be larger when further hypotheses are tested. As noted in Section \ref{sec:online-graph}, closed Alpha-Spending and Online-Graph coincide when $(\gamma_i)_{i\in \mathbb{N}}$ is proportional to a geometric sequence and $g_{j,i}=\gamma_{i-j}$.

\begin{table}[h!]
\centering
\caption{Number of rejections and current significance level $\alpha_{13}$ obtained by different procedures applied on the RECOVERY trial.}
\begin{tabular}[h]{|c|c|c|c|c|c|c|}
\hline
\multirow{2}{*}{\textbf{Procedure}}  & \multicolumn{3}{c|}{\textbf{Number of rejections}} &  \multicolumn{3}{c|}{$\mathbf{\alpha_{13}}$} \\
  & $q=0.6$ & $q=0.7$ & $q=0.8$ & $q=0.6$ & $q=0.7$ & $q=0.8$  \\
  \hline
   Alpha-Spending & $1$ & $2$ & $2$ & $0.00004$ & $0.00021$ & $0.00069$ \\
  Closed Alpha-Spending & $2$ & $2$ & $3$ & $0.00012$ & $0.00042$ & $0.00134$ \\ 
  Online-Graph & $2$ & $2$ & $3$ & $0.00012$ & $0.00042$ & $0.00134$ \\ 
  ADDIS-Spending & $2$ & $3$ & $3$ & $0.00060$ & $0.00112$ & $0.00168$ \\
  Closed ADDIS-Spending & $2$ & $3$ & $3$ & $0.00060$ & $0.00161$ & $0.00210$ \\

  \hline
\end{tabular}
\label{table:results_recovery}
\end{table}

\section{Discussion\label{sec:discussion}}

Contemporary problems, e.g. platform trials and genetics research studies, require control of the FWER in unbounded and sequential multiple testing settings \cite{Retal}. Since the closure principle is fundamental for the construction of multiple testing procedures with FWER control, an extension of the theory is essential. We introduced a novel online closure principle, including a predictability condition for the family of online intersection tests which ensures that the resulting closed testing procedure can be applied in the online setting. Important properties that hold in the classical multiple testing case were transferred to the class of online closed procedures. It was shown that all online procedures with FWER control are also online closed procedures. With this, one can focus on the construction of online closed procedures when aiming for FWER control.  Moreover, we proved that one can restrict to consonant families of intersection tests and provided a  sufficient condition under which the event of rejecting any hypothesis cannot be enlarged. In addition, we showed how short-cuts of online closed procedures can be obtained under consonance. These have a simpler form than in classical multiple testing as the rejection of an intersection hypothesis uniquely determines which individual hypothesis is to be rejected. We used this to derive individual significance levels for short-cuts that are based on $\alpha$-adjustments. With that, new online closed procedures can be derived easily that often improve existing ones which we have demonstrated with the examples of Alpha-Spending and ADDIS-Spending.

 In this paper we focused on the construction of FWER controlling procedures. In online multiple testing, however, many applications aim for a less conservative error rate, e.g. FDR. The reason for this can also be seen when we look at the online procedures that were considered in this paper (e.g. Definition \ref{def:closed-alpha-spending} and \ref{def:closed_addis_spending}). Except for unrealistic extreme cases, the individual significance levels $(\alpha_i)_{i\in \mathbb{N}}$ of all these procedures will tend to $0$ for $i$ to infinity. That is because every additional test increases the probability of committing at least one type I error and thus increases the FWER, whereas rejections may lead to a decrease of the FDR. One could say that both types of procedures have an overall level $\alpha$ available at the beginning of the testing process. But while FWER controlling procedures can only to spend the level on testing, FDR controlling procedures also allow to gain additional significance level from rejections \cite{FS}. Nevertheless, in practice an online multiple testing problem does not necessarily mean that thousands or even millions of hypotheses will be tested, but rather that the number and concrete structure of hypotheses that will be tested is unknown. For example, in platform trials, there may be only a low number of hypotheses to be tested. But since new treatments will be tested over time, online control of the FWER might be required \cite{Retal}. In another approach, that was especially constructed for the modification of machine learning algorithms, the considered online error rate is only controlled over some time window \cite{FES}. In this way, significance level is gained back when hypotheses leave the window, which also makes it reasonable to consider conservative error measures, such as the FWER, in the online setting. A similar approach was considered in \cite{RYWJ}, where the past is ignored in a smooth manner.

 Furthermore, there are approaches to use the closure principle to control other error rates than FWER, such as false discovery proportion (FDP) tail probabilities \cite{goeman2011multiple}. Various types of error rates fall under FDP, such as k-FWER and false discovery exceedance (FDX), and it can even be shown that any admissible FDP procedure must also be a closed procedure \cite{GHS}. In the Appendix, we show that our approach can trivially be extended to obtain an online closure principle for FDP control. However, it is unclear whether the admissibility results derived in \cite{GHS} and the short-cuts derived in \citep{GMKS} still apply in this case. This could be addressed in future work.
 
 There also exist connections between FDR and closed testing. In \cite{RWB}, they introduced an approach to use graphical procedures for FDR control and, in \cite{GMKS}, a connection between Simes-based closed testing and the Benjamini-Hochberg procedure \cite{BH} was shown. In addition, every FDR controlling procedure provides weak FWER control and thereby defines an $\alpha$-level intersection test. Hence, all procedures that were constructed for FDR control can be used to derive new closed testing procedures with FWER or FDP control. This is especially interesting in the online case, as the literature is more advanced for FDR control than for FWER and FDP control.


\newpage

\section*{Appendix}\label{appn}
\subsection*{Online closure principle for FDP control\label{appn:FDP}}

The \textit{false discovery proportion} (FDP) for a some $S\in 2^{\mathbb{N}_f}$, where we denote by $ 2^{\mathbb{N}_f}$ the set of all finite subsets of $\mathbb{N}$, 
 is defined as $$\text{FDP}_{\mathbb{P}}(S)=\frac{|S\cap I_0^{\mathbb{P}}|}{|S|\lor 1} \qquad (\mathbb{P}\in \mathcal{P}).$$
Note that we focus on finite $S\subseteq \mathbb{N}$. On the one hand, $\text{FDP}_{\mathbb{P}}(S)$ is not well-defined for infinite $S\subseteq \mathbb{N}$. On the other hand, from a practical point of view, in most applications we will not have an infinite number of hypotheses at hand, as the infinite testing process is only assumed because we do not know how many hypotheses are to be tested in the future. Therefore, we will only be interested in $\text{FDP}_{\mathbb{P}}(S)$ for finite $S$. Following the notation in \cite{GHS}, we are searching for some random function $\boldsymbol{q}:2^{\mathbb{N}_f}\to [0,1]$ such that for all $\mathbb{P}\in \mathcal{P}$:
$$\mathbb{P}(\boldsymbol{q}(S)\geq \text{FDP}_{\mathbb{P}}(S) \text{ for all } S\in 2^{\mathbb{N}_f}) \geq 1-\alpha.$$
Providing an upper bound for $\text{FDP}(S)$ is equivalent to providing a lower bound for the number of true discoveries $|S\cap I_1|$ \cite{GHS}. Since the number of true discoveries is easier to handle, we are focusing on it in the following. 
A procedure with \textit{true discovery guarantee} is a random function $\boldsymbol{d}:2^{\mathbb{N}_f}\to \mathbb{R}$ such that for all $\mathbb{P} \in \mathcal{P}$:
$$\mathbb{P}(\boldsymbol{d}(S)\leq |S\cap I_1^{\mathbb{P}}| \text{ for all } S\in 2^{\mathbb{N}_f}) \geq 1-\alpha.$$
Furthermore, we call $\boldsymbol{d}$ \textit{online true discovery procedure}, if $\boldsymbol{d}(S)$ is measurable with respect to $\mathcal{F}_{\max(S)}$ for all  $S\in 2^{\mathbb{N}_f}$. Thus, the idea is that  $\boldsymbol{d}(S)$ must be fixed as soon as we can decide on all the individual hypotheses $H_i$ with $i\in S$. The following Theorem is based on the results in \citep{genovese2006exceedance, goeman2011multiple, GHS}.

\begin{theorem}[Online closure principle for true discovery control]
Let a family of $\alpha$-level intersection tests $\boldsymbol{\phi}=(\phi_I)_{I \subseteq \mathbb{N}}$ be given. Then the closed procedure $\boldsymbol{d}^{\boldsymbol{\phi}}$, where $\boldsymbol{d}^{\boldsymbol{\phi}}(S)\coloneqq \min\{|S\setminus I| : I\subseteq \mathbb{N}, \phi_I=0 \}$ for all $S\in 2^{\mathbb{N}_f}$, has true discovery guarantee at level $\alpha$.
In addition, if each $\phi_I$ is an online intersection test and the family of online intersection tests $\boldsymbol{\phi}$ is predictable (Definition \ref{def: predictable family}), then $\boldsymbol{d}^{\boldsymbol{\phi}}$ is an online procedure. 
\end{theorem}

\begin{proof}
   Let $\mathbb{P} \in \mathcal{P}$ be arbitrary. For showing true discovery control of $\boldsymbol{d}^{\boldsymbol{\phi}}$ note that $\boldsymbol{d}^{\boldsymbol{\phi}}(S)>|S\cap I_1^{\mathbb{P}}|$ implies that $\phi_I=1$ for all $I\subseteq \mathbb{N}$ with $|S\setminus I|\leq |S\cap I_1^{\mathbb{P}}|$. Since $|S\setminus I_0^{\mathbb{P}}|= |S\cap I_1^{\mathbb{P}}|$, we especially have $\phi_{I_0^{\mathbb{P}}}=1$. However, this happens with probability at most $\alpha$. Thus, $\mathbb{P}(\boldsymbol{d}^{\boldsymbol{\phi}}(S)\leq |S\cap I_1^{\mathbb{P}}| \text{ for all } S\subseteq I) \geq 1-\alpha$. Furthermore, analogously to Lemma \ref{lemma:closure principle}, we can show that due to the predictability of $\boldsymbol{\phi}$, it holds $\boldsymbol{d}^{\boldsymbol{\phi}}(S)= \min\{|S\setminus I| : I\subseteq \{1,\ldots, \max(S)\}, \phi_I=0 \}$. Since all $\phi_I$ are online intersection tests, $\boldsymbol{d}^{\boldsymbol{\phi}}(S)$ is measurable with respect to $\mathcal{F}_{\max(S)}$.
\end{proof}

\subsection*{Additional simulation results\label{appn:sim}}
\begin{figure}
\centering
\includegraphics[width=12.5cm]{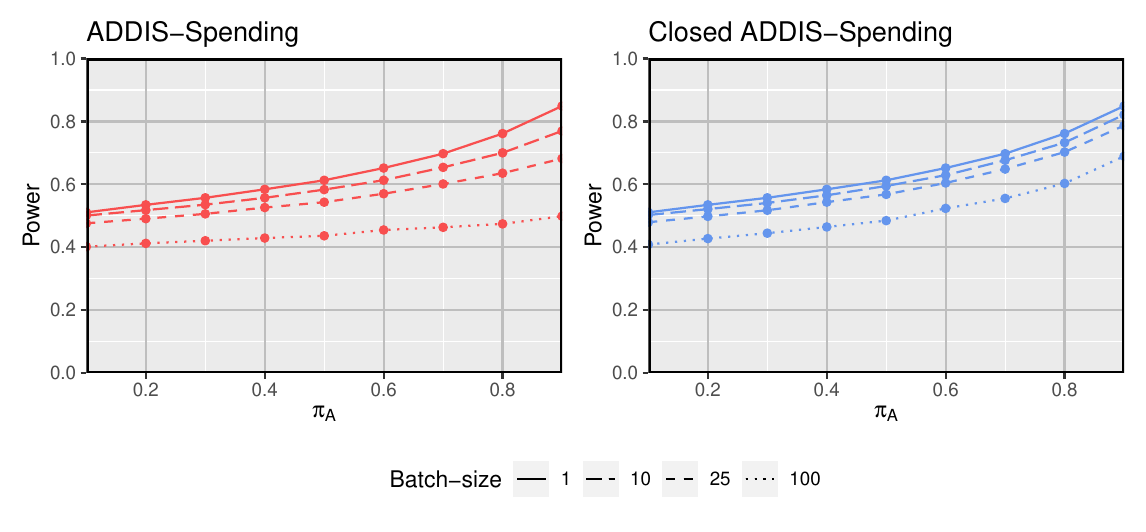}%
\caption{Comparison of ADDIS-Spending and closed ADDIS-Spending in terms of power based on locally dependent $p$-values for different batch-sizes and proportions of false null hypotheses ($\pi_A$); $n=1000$, $\mu_N=-2$, $\mu_A=4$ and $\rho=0.8$ in both plots.\label{fig:simulation_cons}}\end{figure}

\begin{figure}
\centering
\includegraphics[width=12.5cm]{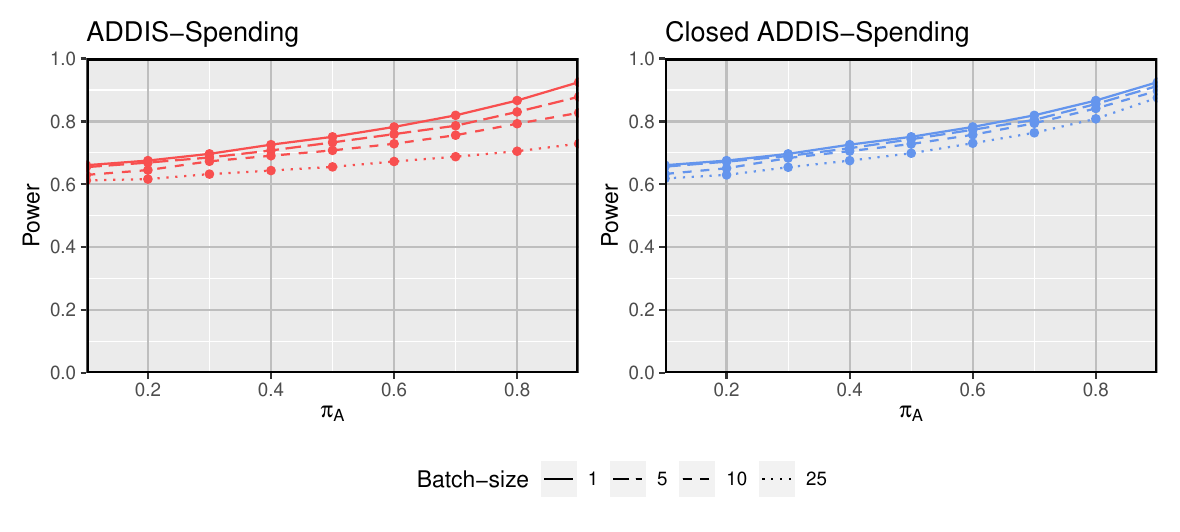}%
\caption{Comparison of ADDIS-Spending and closed ADDIS-Spending in terms of power based on locally dependent $p$-values for different batch-sizes and proportions of false null hypotheses ($\pi_A$); $n=100$, $\mu_N=0$, $\mu_A=4$ and $\rho=0.8$ in both plots.\label{fig:simulation_lown_highmu}}\end{figure}

\begin{figure}
\centering
\includegraphics[width=12.5cm]{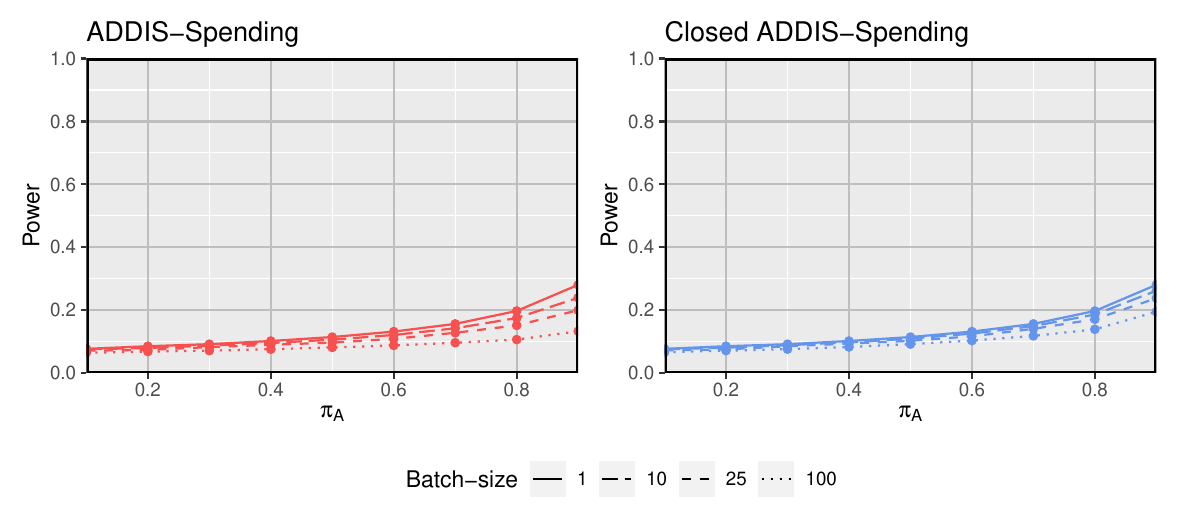}%
\caption{Comparison of ADDIS-Spending and closed ADDIS-Spending in terms of power based on locally dependent $p$-values for different batch-sizes and proportions of false null hypotheses ($\pi_A$); $n=1000$, $\mu_N=0$, $\mu_A=3$ and $\rho=0.8$ in both plots.\label{fig:simulation_highn_lowmu}}\end{figure}

\begin{figure}
\centering
\includegraphics[width=12.5cm]{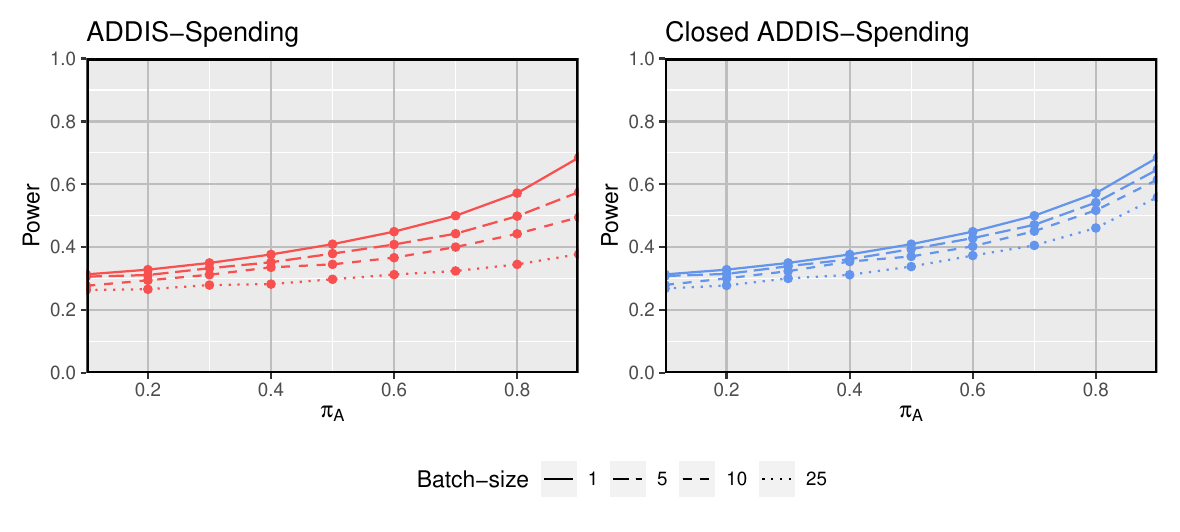}%
\caption{Comparison of ADDIS-Spending and closed ADDIS-Spending in terms of power based on locally dependent $p$-values for different batch-sizes and proportions of false null hypotheses ($\pi_A$); $n=100$, $\mu_N=0$, $\mu_A=3$ and $\rho=0.8$ in both plots.\label{fig:simulation_lown_lowmu}}\end{figure}

In this section we provide additional simulation results based on the design described in Section \ref{subsimsetup}. We applied closed ADDIS-Spending and ADDIS-Spending with the same parameters as in Section \ref{sec:sim_results}. Figure \ref{fig:simulation_cons} shows the results for conservative $p$-values ($\mu_N=-2$). In Figure \ref{fig:simulation_lown_highmu} and Figure \ref{fig:simulation_highn_lowmu}, we reduced the number of hypothesis to $n=100$ and the strength of the alternative to $\mu_A=3$, respectively, and in Figure \ref{fig:simulation_lown_lowmu} we reduced these parameters simultaneously. When we considered the lower number of $n=100$ hypotheses, we also reduced the batch-sizes to $b\in \{1,5,10,25\}$. All plots show the similar behavior that closed ADDIS-Spending is less sensitive than ADDIS-Spending to an increase of the batch-size and thus to locally dependent $p$-values.

\subsection*{Omitted proofs\label{appn:proofs}}

\begin{proof}[Proof of Theorem \ref{theo:consonance procedure}]
We show by induction that $d^{\boldsymbol{\phi}}_i=d^{\boldsymbol{\phi}, s}_i$ for all $i\in \mathbb{N}$.

Initial Case ($i=1$): The predictability of $\boldsymbol{\phi}$ and Lemma \ref{lemma:closure principle} immediately implies that $d^{\boldsymbol{\phi}}_1=1$ if and only if $d^{\boldsymbol{\phi}, s}_1=\phi_{I_1}=1$.

Induction Hypothesis (IH): We assume that $d^{\phi}_j=d^{\phi, s}_j=\phi_{I_j}$ for all $j\leq i-1$, where $i\geq 2$ is arbitrary but fixed.

Induction Step ($i-1\rightarrow i$):  
  \enquote{$\leq$} Since $i\in I_i$, $d^{\boldsymbol{\phi}}_i=1$ immediately implies $d^{\boldsymbol{\phi},s}_i=\phi_{I_i}=1$. \enquote{$\geq$} Assume $d^{\boldsymbol{\phi},s}_i=\phi_{I_i}=1$ and consider an arbitrary subset $J\subseteq \{1,\ldots,i\}$ with $i\in J$. The consonance property implies that there exists a $k\in I_i$ such that $\phi_K=1$ for all $K\subseteq I_i$ with $k\in K$. Since $I_k\subseteq I_i$ for all $k\in I_i$ and $\phi_{I_k}=0$ for all $k\in I_i\setminus \{i\}$, the index satisfying the consonance property has to be $i$. Thus, $H_J$ can be rejected by $\phi_J$ if $J\subseteq I_i$. If $J\not\subseteq I_i$, the definition of $I_i$ ensures that there exists a $j\in J$ with $j<i$ such that $\phi_{I_j}=1$. The induction hypothesis then implies that $H_j$ is rejected by $\boldsymbol{d}^{\boldsymbol{\phi}}$ and hence $H_J$ is rejected by $\phi_J$. Since $J\subseteq \{1,\ldots,i\}$ was arbitrary, all $H_J$ with  $J\subseteq \{1,\ldots,i\}$ and $i\in J$ can be rejected. Moreover, Lemma \ref{lemma:closure principle} implies that $H_i$ is rejected by $\boldsymbol{d}^{\boldsymbol{\phi}}$.
\end{proof}

\begin{proof}[Proof of Theorem \ref{corol:consonance procedure}]
The theorem follows immediately by Theorem \ref{theo:consonance procedure} and Lemma \ref{lemma:short-cut adjustment} below.
\end{proof}

\begin{lemma}\label{lemma:short-cut adjustment}
Assume $(\boldsymbol{\alpha}_I)_{I \subseteq \mathbb{N}}$, where $\boldsymbol{\alpha}_I=(\alpha_i^I)_{i\in I}$, is a predictable family of online sub $\alpha$-adjustments and $(\phi_I)_{I\subseteq N}$ the online intersection tests defined by \eqref{eq:sub_adj}. Then it holds for all $i\in \mathbb{N}$ that $\phi_{I_i}=1$, where $I_i$ is defined in Theorem \ref{theo:consonance procedure}, if and only if $p_i\leq \alpha_i^{I_i}$.
\end{lemma}

\begin{proof}
\enquote{$\Rightarrow$} Assume $\phi_{I_i}=1$. Then there exists a $j\in I_i$ such that $p_j\leq \alpha_j^{I_i}$. Since $I_i=I_j \cup \{k \in \mathbb{N}: j<k<i, \phi_{I_k}=0\} \cup \{i\}$, the predictability of $(\boldsymbol{\alpha}_I)_{I \subseteq \mathbb{N}}$ ensures that $\alpha_j^{I_i}=\alpha_j^{I_j}$ and hence $p_j\leq \alpha_j^{I_j}$. Because $\phi_{I_k}=0$ for all $k\in I_i\setminus \{i\}$, we have $j=i$, meaning $p_i\leq \alpha_i^{I_i}$. \enquote{$\Leftarrow$} $p_i\leq \alpha_i^{I_i}$ implies $\phi_{I_i}=1$ by definition.
\end{proof}

\section*{Acknowledgments}
The authors are grateful for the valuable comments of two anonymous referees and an associate editor, which have led to significant improvements of the paper. In addition, the authors would like to thank Jelle Goeman for a useful discussion.

\section*{Funding} 
L. Fischer acknowledges funding by the Deutsche Forschungsgemeinschaft (DFG, German Research Foundation) – Project number 281474342/GRK2224/2.

M. Bofill Roig is a member of the EU Patient-centric clinical trial platform (EU-PEARL). EU-PEARL has received funding from the Innovative Medicines Initiative 2 Joint Undertaking under grant agreement No 853966. This Joint Undertaking receives support from the European Union's Horizon 2020 research and innovation programme and EFPIA and Children's Tumor Foundation, Global Alliance for TB Drug Development non-profit organization, Spring- works Therapeutics Inc. This publication reflects the author's views. Neither IMI nor the European Union, EFPIA, or any Associated Partners are responsible for any use that may be made of the information contained herein.

\section*{Supplementary Material}
The code for the simulations can be found at the GitHub repository \url{https://github.com/fischer23/Closed-Online-Procedures}.

\bibliographystyle{apalike}  
\bibliography{references.bib}

\end{document}